\documentclass[paper,notoc,nohyper]{JHEP3}

\usepackage{graphicx}
\usepackage{cite}
\newcommand{\text}[1]{#1 }
\def\kt{\tilde{k}}

\def\eps{\epsilon}

\def\Dcz{{\cal D}_0}
\def\Dco{{\cal D}_1}
\def\JET{J}
\def\e{\epsilon}
\def\Flavour{F}
\def\d{\hbox{d}}

\title{Antenna subtraction with hadronic initial states}
\author{A. Daleo\\Institute for Theoretical Physics, ETH, CH-8093 Z\"urich,
Switzerland}
\author{T. Gehrmann, D. Ma\^\i tre\\Institut f\"ur Theoretische Physik, Universit\"at Z\"urich,
Wintherturerstrasse 190,\\CH-8057 Z\"urich, Switzerland}
\abstract{The antenna subtraction method for the computation of higher 
order corrections to jet observables and exclusive cross sections at collider
experiments is extended to include hadronic initial states. In addition to
the already known antenna subtraction with both radiators in the 
final state (final-final antennae), we introduce antenna subtractions with 
one or two radiators in the initial state (initial-final or initial-initial 
antennae). For those, we derive the phase space factorization and
 discuss the allowed phase space mappings at NLO and NNLO. We present  
integrated forms for all antenna functions relevant to NLO calculations, 
and describe the construction of the full antenna subtraction terms at NLO 
on two examples. The extension of the formalism to NNLO is outlined.\\
}
\preprint{ZU-TH 24/06, hep-ph/0612257}

\begin{document}
\section{Introduction}
The calculation of perturbative higher order corrections to 
exclusive observables (especially jet production cross sections, 
but also transverse momentum or rapidity distributions) requires 
a systematic procedure to extract infrared singularities from 
real radiation contributions. These singularities arise if one or 
more final state particles become soft or collinear. For 
the task of next-to-leading 
order (NLO) calculations~\cite{ks} several  systematic and process-independent 
procedures are available. Except for the 
phase space slicing technique~\cite{gieleglover},
all these 
methods~\cite{Catani:1996vz,frixione,ant,cullen,tsnew} 
work by introducing additional terms, that are subtracted from the 
real radiation matrix element at each phase space point. These 
subtraction terms approximate the matrix element in all singular limits, and 
are sufficiently simple to be integrated over part of the phase space
analytically. After this 
integration, infrared divergences 
of the subtraction terms are made explicit, and the integrated subtraction 
terms can be added to the virtual corrections, thus yielding an infrared 
finite result. One of these subtraction methods is antenna 
subtraction~\cite{ant,cullen}, 
which constructs the subtraction terms from so-called antenna functions. 
These antenna functions 
describe all unresolved partonic radiation (soft and collinear) 
between a hard pair of radiator partons.

Extensions to next-to-next-to-leading order (NNLO) are discussed in 
the literature for phase space slicing~\cite{gg}, 
and for several 
subtraction methods~\cite{Kosower:2002su,nnlosub2,nnlosub3,nnlosub4,nnlosub5,Gehrmann-DeRidder:2005cm}. A completely independent approach, avoiding the 
need for analytical integration is the sector decomposition method, which 
has been derived for virtual~\cite{secdecvirt}
 and real radiation~\cite{secdecreal} corrections to NNLO, and applied to 
several observables already~\cite{secdecap}.
Among the subtraction methods, 
only the NNLO formulation of 
the antenna subtraction method~\cite{Gehrmann-DeRidder:2005cm}
 is worked out to a sufficient extent 
to be readily
implemented in the calculation of NNLO corrections to a physical 
process. Using this method, the calculation of $e^+e^- \to 3$~jets
at NNLO accuracy is  currently under way~\cite{our3j}. However, up to now,
the antenna subtraction (both at NLO and NNLO) method was 
developed only to handle unresolved singular radiation off final state 
partons. An extension to radiation off initial state partons was missing 
so far for this method, while all other NLO subtraction methods 
could handle radiation off initial and final state particles.

It is the purpose of this paper to extend the antenna subtraction method 
to include radiation off initial state partons, such that it can be 
used in the calculation of higher order corrections to processes at 
lepton-hadron or hadron-hadron colliders. 
In Section~\ref{sec:ant},
we  recall the structure of antenna subtraction terms and describe their 
systematic derivation. Antenna subtraction for 
radiation off final state partons was formulated already 
in~\cite{cullen,ant,Gehrmann-DeRidder:2005cm} and is 
briefly summarized in Section~\ref{sec:ff}.
In Sections~\ref{sec:if} and~\ref{sec:ii}, we present a detailed 
derivation of antenna subtraction for situations with one or two 
radiator partons in the initial state. In these sections, we provide 
all necessary ingredients for an implementation at NLO, and discuss the 
extension to NNLO. To illustrate the application method, we describe 
two NLO examples in Section~\ref{sec:appl}. Finally, Section~\ref{sec:conc}
contains conclusions and an outlook.

\section{Antenna subtraction}
\label{sec:ant}

To obtain the perturbative corrections to a jet observable at a given  order,
all partonic multiplicity channels contributing to that order  have to be
summed. In general, each partonic channel contains both ultraviolet and
infrared (soft and collinear) singularities.  The ultraviolet poles are removed
by renormalisation in each channel. Collinear poles originating from 
radiation off incoming partons are an inherent feature of the incoming 
partons, and are cancelled by 
redefinition (mass factorization) of the parton distributions. 
The remaining soft and collinear poles cancel  
among each other
only when all partonic channels are summed over.

While infrared singularities from purely virtual corrections are obtained 
immediately after integration over the loop momenta, their extraction is  more
involved for real emission (or mixed real-virtual) contributions. Here, the
infrared singularities only become explicit after integrating  the real
radiation matrix elements over the phase space appropriate to  the jet
observable under consideration. In general, this integration  involves the
(often iterative) definition of the jet observable, such that  an analytic
integration is not feasible (and also not appropriate). Instead,   one would
like to have a flexible method that can be easily adapted to  different jet
observables or jet definitions. Therefore, the infrared singularities  of the real radiation
contributions should be extracted using  infrared subtraction  terms.  The
crucial points that all  subtraction terms must satisfy are  that (a) they
approximate the full  real radiation matrix elements in all singular limits and
(b) are still   sufficiently simple to be integrated analytically over a
section of  phase space that encompasses all regions corresponding to singular
configurations. 

For NLO calculations, several different methods are available to 
derive subtraction terms in a process-independent 
way~\cite{Catani:1996vz,frixione,cullen,ant,tsnew}. 
One of these methods is the so-called 
antenna subtraction~\cite{cullen,ant}. 

In this method, antenna
functions describe the colour-ordered radiation of unresolved
partons between a
pair of hard (radiator) partons. All antenna functions at NLO and NNLO
can be derived systematically from physical matrix elements, as shown
in~\cite{Gehrmann-DeRidder:2005cm,our2j}. 
They can be integrated over the factorized
antenna phase space~\cite{Kosower:2002su} using loop integral reduction
techniques extended to phase space integrals~\cite{ggh}, and then
combined with virtual corrections to partonic processes with lower
multiplicity.

Up to now, antenna subtraction has been formulated at NLO~\cite{cullen,ant}
and NNLO~\cite{Gehrmann-DeRidder:2005cm} only for processes 
with a colourless initial state. In this case, both radiator partons 
are in the final state, we call this situation a final-final antenna. 
For collider observables involving hadronic initial states, there can be 
either one or both partons in the initial state. 
Unresolved radiation off these initial state partons can also be subtracted 
using antenna functions, with one or two radiators in the initial state. We 
call these initial-final and initial-initial antennae. The radiated parton is 
always in the final state. Figures~\ref{fig:ff}--\ref{fig:ii} 
 illustrate 
how a single unresolved parton can be emitted between radiators in the 
final or initial state, and show how all these situations are factorized 
into antenna functions. 
Each antenna contains both collinear limits of the unresolved parton 
with either radiator as well as the soft limit. 
\FIGURE[h!]{
\includegraphics[width=13cm]{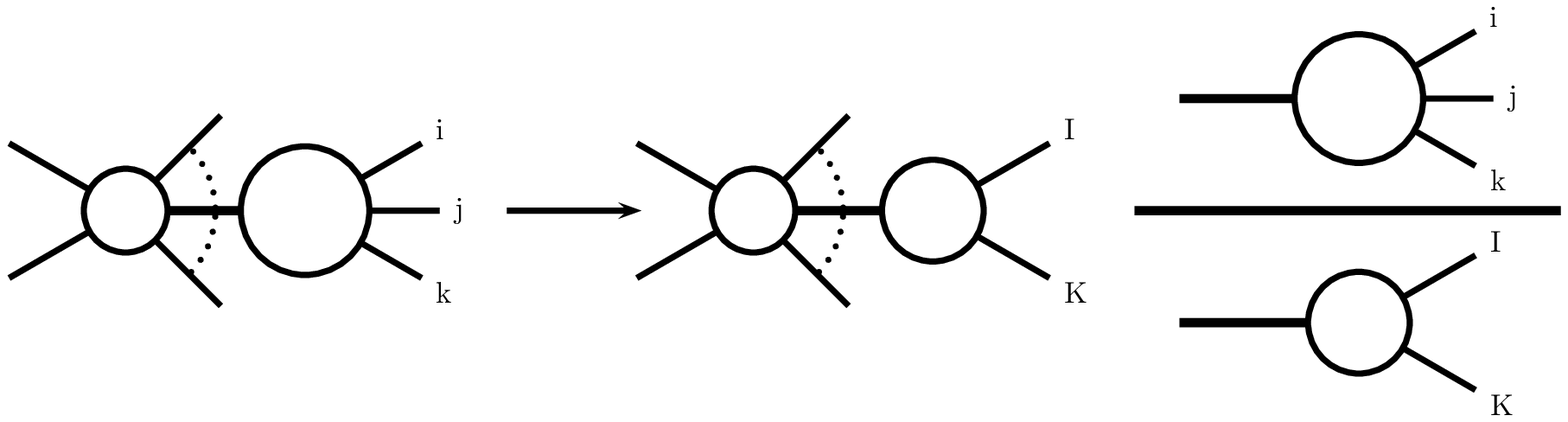}
\label{fig:ff}
\caption{Antenna factorization for the final-final situation.}
}
\FIGURE[h!]{
\includegraphics[width=13cm]{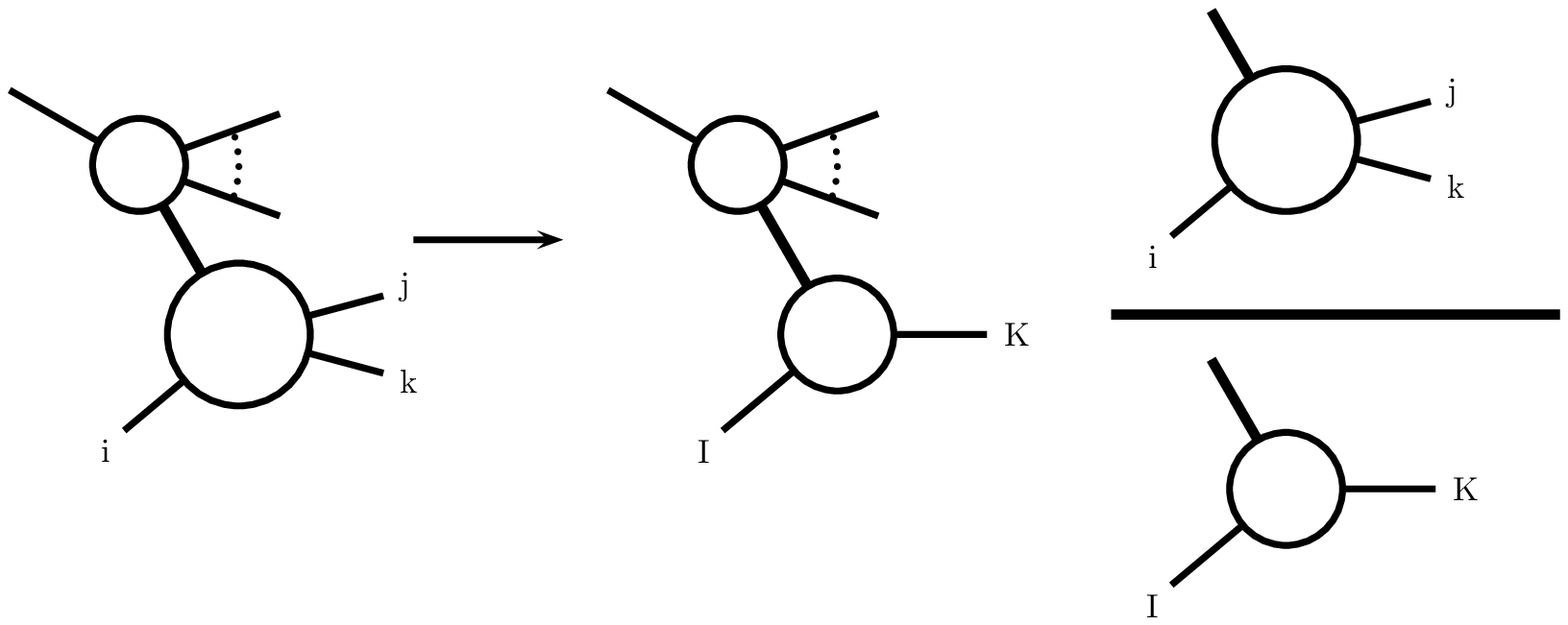}
\label{fig:if}
\caption{Antenna factorization for the initial-final situation.}
}
\FIGURE[h!]{
\includegraphics[width=13cm]{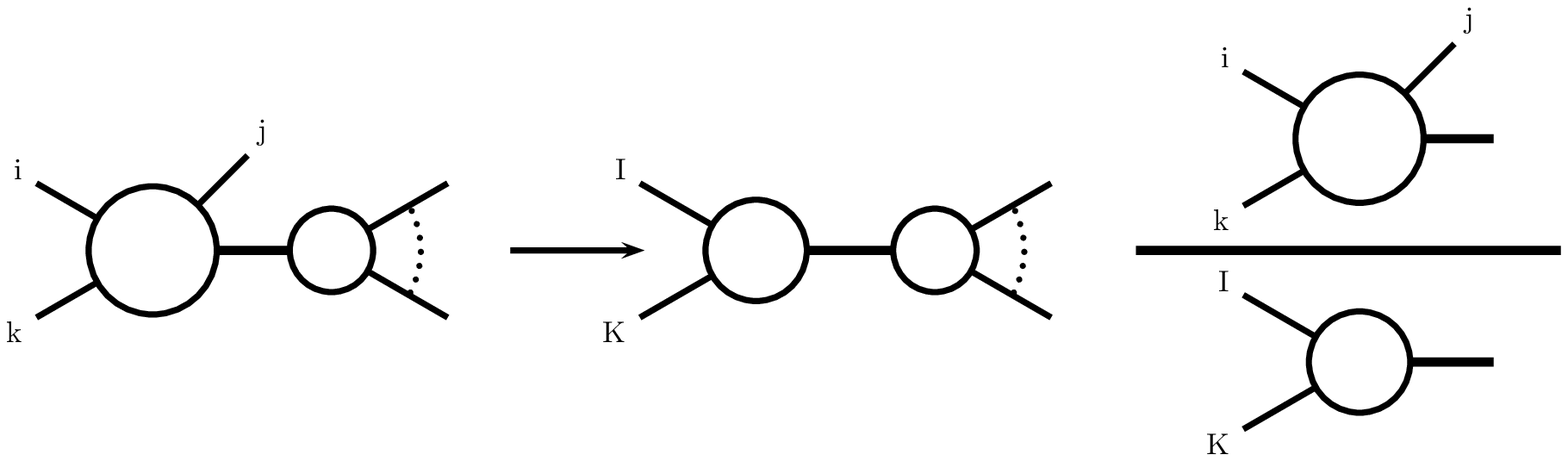}
\label{fig:ii}
\caption{Antenna factorization for the initial-initial situation.}
}

In each situation, the subtraction term is constructed from products of 
antenna functions with reduced matrix elements (with fewer final state 
partons than the original matrix element), and integrated over 
a phase space which is factorized into an antenna phase space (involving all
unresolved partons and the two radiators) multiplied with a reduced phase 
space (where the momenta of radiators and unresolved radiation are replaced 
by two redefined momenta). These redefined momenta can be in the initial state,
if the corresponding radiator momenta were in the initial state. 
The full subtraction term is obtained by summing over all antennae 
required for the problem under consideration. In the most general case
(two partons in the initial state, and two or more hard partons in the final 
state), this sum includes final-final, initial-final and initial-initial 
antennae. 

To specify the notation, we consider the hadronic cross section
\begin{equation}
\d \sigma = \sum_{a,b} \int 
\frac{\d \xi_1}{\xi_1} \frac{\d \xi_2}{\xi_2}\, f_{a/1}(\xi_1) \,f_{b/2}(\xi_2)
\, \d \hat{\sigma}(\xi_1H_1,\xi_2H_2)\;, 
\label{eq:part}
\end{equation}
where $\xi_1$ and $\xi_2$ are the momentum fractions of the 
partons of species $a$ and $b$ in both incoming hadrons, with $f$ being the 
corresponding parton distribution functions and   $H_{1,2}$ denoting the 
incoming hadron momenta. The cases of only one or no incoming hadrons are 
obtained trivially by replacing the relevant $f(\xi)$ by $\delta(1-\xi)$.
The dependence of the parton-level cross section $ \d \hat{\sigma}$ on the 
parton species $a,b$ is obvious, and not stated explicitly for ease of 
notation. It should be noted that the parton level cross section 
$\d \hat\sigma$ is normalized to the hadron-hadron flux factor, 
which is transformed into the parton-parton flux factor by dividing out 
$\xi_1$ and $\xi_2$ in the above.

We define the partonic
tree-level $n$-parton contribution to the $m$-jet cross 
section (for tree-level cross sections $n = m$; we leave $n \neq m$ for later
reference) 
in $d$ dimensions by,
\begin{eqnarray}
{\rm d} \hat\sigma(p_1,p_2) &=&
{\cal N}\,
\sum_{{n}}{\rm d}\Phi_{n}(k_{1},\ldots,k_{n};
p_1,p_2) \nonumber \\ && \times 
\frac{1}{S_{{n}}}\,|{\cal M}_{n}(k_{1},\ldots,k_{n};p_1,p_2)|^{2}\; 
\JET_{m}^{(n)}(k_{1},\ldots,k_{n}).
\label{eq:sigm}
\end{eqnarray}
The definition of the observable $\JET_{m}^{(n)}(k_{1},\ldots,k_{n})$
can depend on $H_{1,2}$ (for example through cuts on the 
jet rapidities), although this is not stated explicitly here. 
The normalization factor 
${\cal N}$ includes all QCD-independent factors as well as the 
dependence on the renormalized QCD coupling constant $\alpha_s$,
$\sum_{n}$ denotes the sum over all configurations with $n$ partons,
${\rm d}\Phi_{n}$ is the phase space for an $n$-parton final state with total
four-momentum $p_1^{\mu}+p_2^{\mu}$ in
$d=4-2\e$ space-time dimensions,
\begin{equation}
\d \Phi_n(k_{1},\ldots,k_{n};p_1,p_2) 
= \frac{\d^{d-1} k_1}{2E_1 (2\pi)^{d-1}}\; \ldots \;
\frac{\d^{d-1} k_n}{2E_n (2\pi)^{d-1}}\; (2\pi)^{d} \;
\delta^d (p_1+p_2 - k_1 - \ldots - k_n) \,,
\end{equation}
while $S_{n}$ is a
symmetry factor for identical partons in the final state.
$|{\cal M}_{n}|^2$ denotes a 
squared, colour-ordered tree-level $n$-parton matrix element.

Antenna subtraction terms $\d \sigma^S$ are constructed 
using parton-level antenna subtraction terms $\d \hat\sigma^S$ as in 
(\ref{eq:part}), such that 
\begin{displaymath}
{\rm d} \sigma -  {\rm d} \sigma^{S}
\end{displaymath}
is finite in all unresolved limits, and that phase space integrals 
contained in it can
be carried out numerically. 

In the following, we will briefly summarize the features of 
antenna subtraction in the final-final case, and derive the antenna 
phase spaces and antenna functions for the two other cases.

\section{Final-final configurations}
\label{sec:ff}
In configurations involving final-final antennae, both radiators are in the final state. 
This case was described previously in detail at NLO~\cite{cullen,ant} and 
NNLO~\cite{Gehrmann-DeRidder:2005cm}. The NLO subtraction term for 
an unresolved 
parton $j$, emitted between hard final-state 
radiators $i$ and $k$  is depicted in 
Figure~\ref{fig:ff}. It reads
\begin{eqnarray}
\lefteqn{{\rm d}\hat\sigma^{S,(ff)}(p_1,p_2) =
{\cal N}\,\sum_{m+1}{\rm d}\Phi_{m+1}(k_{1},\ldots,k_{m+1};p_1,p_2)
\frac{1}{S_{{m+1}}} }\nonumber \\ 
&\times &\sum_{j}\;X^0_{ijk}\,
|{\cal M}_{m}(k_{1},\ldots,{K}_{I},{K}_{K},\ldots,k_{m+1};
p_1,p_2)|^2\,
\JET_{m}^{(m)}(k_{1},\ldots,{K}_{I},{K}_{K},\ldots,k_{m+1})
\;\nonumber \\
\label{eq:sub1}
\end{eqnarray}
The subtraction term involves the $m$-parton amplitude 
depending only on the redefined on-shell momenta
$k_{1},\ldots,{K}_{I},{K}_{K},\ldots,k_{m+1}$ 
where ${K}_{I},{K}_{K}$
 are linear combinations of $k_{i},k_{j},{k}_{k}$
 while the tree antenna function $X^0_{ijk}$ depends only on 
$k_{i},k_{j},{k}_{k}$.
$X^0_{ijk}$ describes all of the configurations (for this colour-ordered
amplitude)
where parton $j$ is unresolved.

The jet function $\JET^{(m)}_m$ in (\ref{eq:sub1}) does not depend on the 
individual momenta ${k}_{i}$, $k_j$ and ${k}_{k}$, but only on 
${K}_{I},{K}_{K}$. One can therefore carry out the integration
over the unresolved dipole phase space appropriate to 
   ${k}_{i}$, $k_j$ and ${k}_{k}$ analytically, exploiting the 
factorization of the phase space,
\begin{eqnarray}
\label{eq:psx3}
\lefteqn{\d \Phi_{m+1}(k_{1},\ldots,k_{m+1};p_1,p_2)  = }
\nonumber \\ &&
\d \Phi_{m}(k_{1},\ldots,{K}_{I},{K}_{K},\ldots,k_{m+1};p_1,p_2)
\cdot 
\d \Phi_{X_{ijk}} (k_i,k_j,k_k;{K}_{I}+{K}_{K},0)\;.
\end{eqnarray}
The NLO antenna phase space $\d \Phi_{X_{ijk}}$ 
is proportional to the three-particle phase space relevant to a $1\to 3$ 
decay.

At NNLO, one has to consider the emission of one parton in a one-loop 
corrected process, or the emission of two partons at tree level. Both these 
were described in detail in~\cite{Gehrmann-DeRidder:2005cm}. 
While the one-loop antenna subtraction is largely an extension of the 
above with the replacement of the tree-level antenna function 
by a one-loop antenna function, $X^0\to X^1$, several new features 
appear in the subtraction of two unresolved partons at tree-level. 

In particular, one must pay attention to the colour-connection of the 
two unresolved partons. If they are colour-unconnected or almost 
colour-unconnected (sharing a common radiator), the subtraction term is 
obtained by iterating the procedure employed at NLO, now yielding 
products of two antenna functions. If both unresolved partons $j,k$ are 
colour-connected, new four-parton antenna functions $X_{ijkl}$ 
appear in the subtraction terms:
\begin{eqnarray*}
\lefteqn{{\cal N}\,\sum_{m+2}{\rm d}\Phi_{m+2}(k_{1},\ldots,k_{m+2};p_1,p_2)
\frac{1}{S_{{m+2}}}  \sum_{jk}\; X^0_{ijkl}}
\nonumber \\
&\times&
|{\cal M}_{m}(k_{1},\ldots,{K}_{I},{K}_{L},\ldots,k_{m+2};p_1,p_2)|^2\,
\JET_{m}^{(m)}(k_{1},\ldots,{K}_{I},{K}_{L},\ldots,k_{m+2})\;
\Bigg]\;,
 \end{eqnarray*}
where the sum runs over all colour-adjacent pairs $j,k$ and implies the 
appropriate selection of hard momenta $i,l$.
As before, the subtraction term involves the $m$-parton amplitude 
evaluated with on-shell momenta $k_{1},\ldots,{K}_{I},{K}_{L},\ldots,k_{m+2}$
where now ${K}_{I}$ and ${K}_{L}$ are a linear combination of
$k_i$, $k_j$, $k_k$ and $k_l$.  
As for the NLO antenna of the previous section, 
the tree antenna function $X^0_{ijkl}$
depends only on $k_i,k_j,k_k,k_l$.  Particles $i$ and $l$ play the
 role of the
radiators while $j$ and $k$ are the radiated partons.  

Once again, the jet function $\JET^{(m)}_m$ in the above equation
depends only on the parent momenta 
${K}_{I},{K}_{L}$ and not $k_i,\ldots,k_l$. 
One can therefore carry out the integration
over the unresolved antenna phase space (or part thereof)
analytically, exploiting the 
factorization of the phase space,
\begin{eqnarray}
\label{eq:psx4}
\lefteqn{\d \Phi_{m+2}(k_{1},\ldots,k_{m+2};p_1,p_2)  = }\nonumber \\
&&\d \Phi_{m}(k_{1},\ldots,{K}_{I},{K}_{L},\ldots,k_{m+1};p_1,p_2)
\cdot 
\d \Phi_{X_{ijkl}} (k_i,k_j,k_k,k_l;{K}_{I}+{K}_{L},0)\;.
\end{eqnarray}
This phase space factorization must be carried out such that all 
unresolved limits are reproduced correctly. The most general 
parameterization for this case is derived in \cite{Kosower:2002su}.
It should be noted that $\d \Phi_{X_{ijkl}}$ is proportional to the 
$1\to 4$ parton phase space; the analytical integration of the 
antenna functions over this phase space can thus be carried out with 
standard methods~\cite{cut,ggh}.

\section{Initial-final configurations}
\label{sec:if}
In the presence of hadrons in the initial state, matrix elements
exhibit singularities that are not accounted by the subtraction
terms discussed in the previous section. These singularities are 
due to soft or collinear radiation within an antenna where one or 
the two hard partons are in the initial state. 

As discussed in \cite{Gehrmann-DeRidder:2005cm}, the
terms necessary to subtract singularities associated with colored 
particles in the initial state can be simply obtained 
by crossing the corresponding antennae for final state singularities.
Due to the different kinematics involved, the factorization of phase space 
is slightly more involved and the corresponding phase space mappings are 
different. To cancel explicit infrared poles in virtual contributions and
in terms arising from 
parton distribution 
mass factorization, the crossed antennae must be
integrated, analytically, over the corresponding phase space.
In this section we will present the antennae and phase space mappings to subtract 
singularities when only one of the radiating partons is in the initial state. 

\subsection{Subtraction terms for initial-final singularities}
Subtraction terms in the case of one hard parton in the initial state
are built in the same fashion as for the final-final case (formula (2.5) 
in \cite{Gehrmann-DeRidder:2005cm}). 
We have the following subtraction term
associated to a hard radiator parton $i$ with momentum $p$ in the initial state:
\begin{eqnarray}\label{eq:subif}
&&\d\hat{\sigma}^{S,(if)}(p,r)={\cal N}\sum_{m+1}\d\Phi_{m+1}(k_1,\dots,k_{m+1};p,r)
  \,\frac{1}{S_{m+1}}\nonumber\\
&&\times\sum_{j} X^{0}_{i,jk}
  \left|{\cal M}_m(k_1,\dots,K_{K},\dots,k_{m+1};x p,r)\right|^2\,
  J^{(m)}_{m}(k_1,\dots,K_{K},\dots,k_{m+1})\,.
\end{eqnarray}
The additional momentum $r$ stands for the momentum of the second 
incoming particle, for example,
a virtual boson in DIS, or a second incoming parton in a hadronic
collision process. 
This contribution has to be appropriately convoluted with
the parton distribution function $f_i$.   
The tree antenna $X^{0}_{i,jk}$, depending only on 
the original momenta $p$, $k_j$ and $k_k$, contains all the
configurations in which parton $j$ becomes unresolved. 
The $m$-parton amplitude depends only on redefined on-shell momenta 
$k_1,\dots,K_{K},\dots,$ and on the momentum fraction $x$. 
 In the case where the second incoming particle is a parton, there is an additional convolution
with the 
parton distribution of parton $r$ and corresponding subtraction terms
associated with it.

The jet function, $J^{(m)}_{m}$, in (\ref{eq:subif}) depends on the
momenta $k_j$ and $k_k$ only through $K_K$. Thus, provided a
suitable factorization of the phase space, one can perform the 
integration of the antennae analytically. Due to the hard particle in the
initial state, the factorization of phase space is not as straightforward 
as for final-final antennae. We start from the $(m+1)$-particle phase space
\begin{equation}
\d\Phi_{m+1}(k_1,\dots,k_{m+1};p,r)=(2\pi)^{d}\delta\left(r+p-\sum_l k_l\right)\,\prod_{l}[\d k_l]\,
\end{equation}
where $[\d k]=\d^dk\,\delta^{+}(k^2)/(2\pi)^{d-1}$. We insert
\begin{equation}\label{eq:ifpm1}
1=\int \d^d q\,\delta\left(q+p-k_j-k_k\right)\,,
\end{equation}
and
\begin{equation}\label{eq:ifpm2}
1=\frac{Q^2}{2\pi}\int\frac{\d x}{x}\int [\d K_{K}]\,(2\pi)^{d}\delta\left(q+x p-K_{K}\right)\,,
\end{equation}
with $Q^2=-q^2$. Finally, integrating over $q$, the phase space can be factorized in an $m$-parton phase space
convoluted with a two particle phase space:
\begin{eqnarray}
\d\Phi_{m+1}(k_1,\dots,k_{m+1};p,r)&=&\d\Phi_{m}(k_1,\dots,K_{K},\dots,k_{m+1};x p,r)\nonumber\\
&\times&\frac{Q^2}{2\pi}\d\Phi_{2}(k_j,k_k;p,q)\frac{\d x}{x}\,.
\end{eqnarray}
Replacing the phase space in (\ref{eq:subif}), we can explicitly carry out the integration of the
antenna factors over the two particle phase space. When combining the integrated subtraction terms
with virtual contributions and mass factorization terms, it turns 
out to be convenient to normalize the 
integrated antennae as follows 
\begin{equation}\label{eq:aint}
{\cal X}_{i,jk}=\frac{1}{C(\epsilon)}\int \d\Phi_2 \frac{Q^2}{2\pi} X_{i,jk}\,,
\end{equation}
where
\begin{equation}
C(\epsilon)=\left({4\pi}\right)^{\epsilon}\,\frac{e^{-\epsilon\gamma_E}}{8\pi^2}\,.
\end{equation}
The integrated form of the subtraction term is then
\begin{eqnarray}
\d\hat{\sigma}^{S,(if)}(p,r)&=&\sum_{m+1}\sum_{j}
\frac{{\cal N}}{S_{m+1}}\int\frac{dx}{x}\,
C(\epsilon)\,{\cal X}_{i,jk}(x)\,\d\Phi_{m}(k_1,\dots,K_{K},\dots,k_{m+1};x\,p,r)\nonumber\\
&&\times\left|{\cal M}_m(k_1,\dots,K_{K},\dots,k_{m+1};x\,p,r)\right|^2
J^{(m)}_{m}(k_1,\dots,K_{K},\dots,k_{m+1})\,.
\end{eqnarray}
Finally, the subtraction term has to be convoluted with the parton
distribution functions to give the corresponding contribution to the
hadronic cross section. The explicit poles in the integrated form
cancel the corresponding ones in the virtual and PDFs mass factorization 
contributions. To carry out the explicit cancellation of poles, it is 
convenient to recast, by a simple change of variables, the integrated 
subtraction term, once convoluted with the PDFs, in the following form
\begin{eqnarray}
d{\sigma}^{S,(if)}(p,r)&=&\sum_{m+1}\sum_{j}
\frac{S_{m}}{S_{m+1}}\int\frac{\d\xi_1}{\xi_1}\int\frac{\d\xi_2}{\xi_2}\int_{\xi_1}^{1}\frac{\d x}{x}
f_{i/1}\left(\frac{\xi_1}{x}\right)\,f_{b/2}\left(\xi_2\right)\nonumber\\
&&\times C(\epsilon)\,{\cal X}_{i,jk}(x)\,\d\hat{\sigma}^B(\xi_1 H_1,\xi_2 H_2)\,.
\end{eqnarray}
This convolution has already the appropriate 
structure and mass factorization can be carried out explicitly leaving a finite 
contribution. The remaining phase space integration, implicit in
the Born cross section, $\d\hat{\sigma}^B$, and the convolutions can 
be safely done numerically. When considering reactions with only one 
incoming hadron, the second PDF has to be replaced by a Dirac delta. Reactions 
with two hadrons will require additional subtractions containing initial-final 
antennae involving the second parton in the initial state and initial-initial
antennae as well. This case will be discussed in Section \ref{sec:ii} below.
  
\subsection{Phase space mapping}

Proper subtraction of infrared singularities requires that the momenta mapping 
satisfies
\begin{eqnarray}
&&x p\rightarrow p\,,\qquad K_{K}\rightarrow k_k\qquad\mbox{when $j$ becomes soft}\,,\nonumber\\
&&x p\rightarrow p\,,\qquad K_{K}\rightarrow k_j+k_k
\qquad\mbox{when $j$ becomes collinear with $k$}\,,\nonumber\\
&&x p\rightarrow p-k_j\,,\qquad K_{K}\rightarrow k_k
\qquad\mbox{when $j$ becomes collinear with $i$}\,.
\end{eqnarray}
In this way, infrared singularities are subtracted locally, except for 
angular correlations, {\em before convoluting with the
parton distributions}. That is, 
matrix elements and subtraction terms are convoluted together with PDFs.
In addition, the redefined momentum, $K_{K}$, must be on shell and 
momentum must be conserved, $p-k_j-k_k=x p-K_{K}$, for the phase space to 
factorize as above.

As discussed, in the case of configurations with two hard radiators in 
the final state, the three-to-two-parton map of \cite{Kosower:2002su} is suitable, 
as it treats both collinear limits symmetrically and there is only one mapping describing
all the singular configurations contained in the antennae. 

When subtracting initial state singularities, however, the 
mapping of \cite{Kosower:2002su} leads to a non factorizing phase space.
The decisive point is that this mapping, 
modified to account for a particle in the initial state, introduces a 
new initial state momentum, $P$ as a linear combination of 
$p$, $k_j$ and $k_k$. However, as there is no integration over $P$,
the factorization of phase space would not be complete, because the 
$m$-parton matrix elements depend on $P$. On the other hand, if $P$ is proportional to 
$p$, factorization is granted, in the form of a 
convolution between the reduced matrix elements and the integrated
antennae, as we detailed above. In this case, we inmediately obtain 
the dipole momentum mappings of~\cite{Catani:1996vz}, 
combined into a single mapping interpolating between all the singular
limits of the antennae. Explicitly:
\begin{eqnarray}\label{eq:ifmappingNLO}
x&=&\frac{s_{1j}+s_{1k}-s_{jk}}{s_{1j}+s_{1k}}\,,\nonumber \\
K_{K}&=&k_j+k_k-(1-x)p\,,
\end{eqnarray}
where $s_{1j}=(p-k_j)^2$, etc. If parton $j$ becomes soft or collinear to
parton $k$, $x\rightarrow 1$. If parton $j$ becomes collinear with the
initial state parton $i$, $x=1-z$ with $z$ the fraction of $p$ carried
by $j$. 

The mapping in eq. (\ref{eq:ifmappingNLO}) is, in addition, easily generalized
to deal with more than one parton becoming unresolved. 
The building blocks for the double real radiation in the initial-final 
situation are colour-ordered
four-parton antenna functions $X_{i,jkl}$, with 
one radiator parton $i$ (with momentum $p$)
in the initial state, two unresolved partons $j,k$
and one radiator parton $l$ in the final state. Starting with the
generalization of (\ref{eq:ifpm1}) to three particles in the final state,
and combining with (\ref{eq:ifpm2}) we have the following mapping at NNLO: 
\begin{eqnarray}\label{eq:ifmappingNNLO}
x&=&\frac{s_{1j}+s_{1k}+s_{1l}-s_{jk}-s_{jl}-s_{kl}}{s_{1j}+s_{1k}+s_{1l}}\,,
\nonumber \\
K_{L}&=&k_j+k_k+k_l-(1-x)p\,,
\end{eqnarray}
where $k_j$, $k_k$ and $k_l$ are the three final state momenta involved in 
the subtraction term. This mapping can be 
obtained from the tripole mapping~\cite{ggh,uwer}
 for final-final configurations at NNLO.
It satisfies the appropriate limits in all double singular configurations:
\begin{enumerate}
\item $j$ and $k$ soft: $x\rightarrow 1$, $K_L\rightarrow k_l$,
\item $j$ soft and $k_k\parallel k_l$: $x\rightarrow 1$, $K_L\rightarrow k_k+k_l$,
\item $k_j=zp\parallel p$ and $k$ soft: $x\rightarrow 1-z$, $K_L\rightarrow
  k_l$,
\item $k_j=zp\parallel p$ and $k_k\parallel k_l$: $x\rightarrow 1-z$, $K_L\rightarrow
  k_k+k_l$,
\item $k_j\parallel k_k\parallel k_l$: $x\rightarrow 1$, $K_L\rightarrow
  k_j+k_k+k_l$,
\item $k_j+k_k=zp\parallel p$: $x\rightarrow 1-z$, $K_L\rightarrow
  k_l$,
\end{enumerate}  
where partons $j$ and $k$ can be interchanged in all the cases.

The construction of NNLO antenna subtraction terms requires moreover that
all single unresolved limits of the four-parton antenna function
$X_{i,jkl}$
have to be subtracted, such that the resulting subtraction term is active only
in its double unresolved limits. A systematic subtraction of these single 
unresolved limits by products of two three-parton antenna functions can be 
performed only if the NNLO phase space mapping turns into an NLO phase space 
mapping in its single unresolved limits~\cite{Kosower:2002su}.

In the limits where parton $j$ becomes unresolved, we denote the parameters 
of the reduced NLO phase space mapping 
(\ref{eq:ifmappingNLO}) by ${x}^\prime$ and $K_L^\prime$. 
We find for 
(\ref{eq:ifmappingNNLO}):
\begin{enumerate}
\item $j$ becomes soft: $$
x \to 
\frac{s_{1k}+s_{1l}-s_{kl}}{s_{1k}+s_{1l}} = {x}^\prime \,,\qquad 
K_{L} \to k_k+k_l-(1-x)p = K_L^\prime \,. $$
\item $k_j\parallel k_k$, $k_j+k_k = K_K$: $$
x \to  
\frac{s_{1K}+s_{1l}-s_{Kl}}{s_{1K}+s_{1l}} = {x}^\prime \,,\qquad 
K_{L} \to k_K+k_l-(1-x)p = K_L^\prime\,. $$
\item $k_j = zp \parallel p$: $$
x \to \frac{(1-z)(s_{1k}+s_{1l})-s_{kl}}{s_{1k}+s_{1l}} = (1-z) x^\prime 
\,,\qquad K_{L} \to k_k+k_l-(1-x^\prime)(1-z) p = K_L^\prime.
$$
\end{enumerate}
It can be seen that in the first two limits, the NLO mapping 
involves the original incoming momentum $p$, while in the last limit
(initial state collinear emission), it involves the rescaled incoming 
momentum $(1-z)p$. 
To subtract all three single 
unresolved limits of parton $j$ between emitter partons $i$ and $k$ 
from $X_{i,jkl}$, one needs to subtract from it the 
product of two three-parton 
antenna functions $X_{i,jk}\cdot X_{I,Kl}$. The phase space mapping 
relevant to these terms is the iteration of two NLO phase space
mappings. Analytical integration of these terms with this mapping 
will result in a double convolution of both antenna functions with 
the reduced matrix element. 

Equally, parton $k$ can become unresolved. Expressing the reduced NLO 
phase space mapping by  ${x}^{\prime\prime}$ and $K_L^{\prime\prime}$. 
We find for 
(\ref{eq:ifmappingNNLO}):
\begin{enumerate}
\item $k$ becomes soft: $$
x \to 
\frac{s_{1j}+s_{1l}-s_{jl}}{s_{1j}+s_{1l}} = {x}^{\prime\prime} \,,\qquad 
K_{L} \to k_j+k_l-(1-x)p = K_L^{\prime\prime} \,. $$
\item $k_k\parallel k_j$, $k_j+k_k = K_K$: $$
x \to  
\frac{s_{1K}+s_{1l}-s_{Kl}}{s_{1K}+s_{1l}} = {x}^{\prime\prime} \,,\qquad 
K_{L} \to k_K+k_l-(1-x)p = K_L^{\prime\prime}\,. $$
\item $k_k\parallel k_l$, $k_l+k_k = K_K$: $$
x \to  
\frac{s_{1K}+s_{1j}-s_{Kj}}{s_{1K}+s_{1j}} = {x}^{\prime\prime} \,,\qquad 
K_{L} \to k_K+k_j-(1-x)p = K_L^{\prime\prime}\,. $$
\end{enumerate}
In all limits, the reduced NLO mapping involves the original incoming momentum 
$p$. Consequently, the three single 
unresolved limits of parton $k$ between emitter partons $j$ and $l$ 
can be subtracted from 
$X_{i,jkl}$ 
by a product of a final-final and an initial-final three-parton 
antenna function $X_{jkl}\cdot X_{i,JL}$. The phase space mapping 
relevant to these terms is the product of an NLO final-final phase space 
mapping with an initial-final mapping. Integration of the 
 final-final antenna phase space yields an constant, not involving an extra 
convolution, such that these terms appear in the integrated subtraction term 
only with a single convolution with the reduced matrix element.

\subsection{NLO antenna functions}\label{sec:antenna-if}
We now present explicit results for all the antenna functions necessary 
to subtract infrared singularities associated with one particle in the
initial state. The unintegrated form of all of them can be obtained from 
the corresponding expressions for the tree level three particles antennae
in~\cite{Gehrmann-DeRidder:2005cm} by appropriate crossing of particles 
from the final to the initial state. In the cases where there are different 
particles in the final state, there are more than one possible crossing and,
thus, more than one corresponding antenna.

The invariants for antenna $X_{i,jk}$ are defined as ~$s=(k_j+k_k)^2$, ~$t=(p-k_j)^2$,
~$u=(p-k_k)^2$ and $Q^2=-q^2$, where $q=p-k_j-k_k$. For the integrated antennae 
we define $x=Q^2/(2\,p\cdot q)$.
The colour-ordered splitting kernels are given by
\begin{eqnarray}
&&p^{(0)}_{qq}(x)=\frac{3}{2}\,\delta(1-x)+2\Dcz(x)-1-x\,,
\nonumber \\
&&p^{(0)}_{qg}(x)=1-2x+2x^2\,,\nonumber \\
&&p^{(0)}_{gq}(x)=\frac{2}{x}-2+x\,,\nonumber \\
&&p^{(0)}_{gg}(x)=\frac{11}{6}\,\delta(1-x)+2\Dcz(x)+\frac{2}{x}-4+2x-2x^2\,,
\nonumber \\
&&p^{(0)}_{gg,F}(x)=-\frac{1}{3}\,\delta(1-x)\,,
\end{eqnarray}
where we introduced the distributions
\begin{displaymath}
{\cal D}_n(x) = \left(\frac{\ln^n(1-x)}{1-x}\right)_+\;.
\end{displaymath}
The colour-ordered infrared singularity operators are
 as in~\cite{Gehrmann-DeRidder:2005cm}:
\begin{eqnarray}
{\bf I}^{(1)}_{q\bar q} (\e,s_{q\bar q})
&=& - \frac{e^{\e \gamma}}{2\Gamma(1-\e)}\, \left[
\frac{1}{\e^2}+\frac{3}{2\e} \right] \, \Re(-s_{q\bar q})^{-\e} \;,
\nonumber \\ 
{\bf I}^{(1)}_{q g}(\e,s_{q g})
 &=& - \frac{e^{\e \gamma}}{2\Gamma(1-\e)}\, \left[
\frac{1}{\e^2}+\frac{5}{3\e} \right] \, \Re(-s_{qg})^{-\e}  \;,
\nonumber\\
{\bf I}^{(1)}_{gg}(\e,s_{gg})
 &=& - \frac{e^{\e \gamma}}{2\Gamma(1-\e)}\, \left[
\frac{1}{\e^2}+\frac{11}{6\e} \right]\, \Re(-s_{gg})^{-\e} \;,
\nonumber\\ 
{\bf I}^{(1)}_{q\bar q,\Flavour}(\e,s_{q\bar q}) &=& 0  \;,
\nonumber\\ 
{\bf I}^{(1)}_{q g,\Flavour}(\e,s_{q g}) &=&  \frac{e^{\e \gamma}}{2\Gamma(1-\e)}\, 
\frac{1}{6\e} \, \Re(-s_{qg})^{-\e}\;, \nonumber \\
{\bf I}^{(1)}_{g g,\Flavour}(\e,s_{gg}) &=&  \frac{e^{\e \gamma}}{2\Gamma(1-\e)}\, 
\frac{1}{3\e} \, \Re(-s_{gg})^{-\e} \;.
\label{eq:Ione}
\end{eqnarray}
Although the antenna functions are obtained by simple crossing of the 
antenna functions for the final-final case, there are some important 
differences in the decomposition of antenna functions into sub-antennae. 
In the final-final case, antenna functions involving a hard gluon radiating 
unresolved gluons had to be split into different configurations since 
any final state gluon could be identified as the hard radiator. This ambiguity is 
no longer present if a gluon is crossed into the initial state, since an 
initial state gluon is hard by kinematical constraints. Instead, a 
different ambiguity appears, since the initial state gluon can split either 
into a quark or into a gluon, thus leading to two possible reduced matrix 
elements. This ambiguity requires decomposition of the relevant 
gluon-initiated antenna functions  into sub-antennae according to criteria 
completely different from the final-final situation, as will be 
discussed in Section~\ref{sec:gini} below. 

\subsubsection{Quark initiated antennae}
We consider first antennae with a quark in the initial state. There is one quark-quark 
antenna, given by
\begin{equation}
A^{0}_{q,gq}=-\frac{1}{Q^2}\left(\frac{2u}{s}+\frac{2u}{t}+\frac{2u^2}{st}+
        \frac{t}{s}+\frac{s}{t}\right)+{\cal O}(\epsilon)\,.
\end{equation}
Its integral over the phase space (normalized as in eq. (\ref{eq:aint}))
gives
\begin{eqnarray}
{\cal A}^{0}_{q,gq}&=&-2{\bf I}^{(1)}_{q\bar{q}}(Q^2)\,\delta(1-x)+(Q^2)^{-\eps}\left[
-\frac{1}{2\eps}p^{(0)}_{qq}(x)+
\left(\frac{7}{4}-\frac{\pi^2}{6}\right)\delta(1-x)\right.\nonumber\\
&&-\left.\frac{3}{4}\Dcz(x)+\Dco(x)
-\frac{3-x}{2}-\frac{1+x}{2}\log(1-x)-\frac{1+x^2}{2(1-x)}\log(x)+{\cal O}(\eps)\right]\,.
\end{eqnarray}
There are three quark-gluon antennae given by
\begin{eqnarray}
D^{0}_{q,gg}&=&\frac{1}{(Q^2)^2}\left(\frac{s^2}{t}+\frac{s^2}{u}+\frac{t^2}{s}+\frac{4t^2}{u}+
\frac{4u^2}{s}+\frac{4u^2}{t}+\frac{3st}{u}+\frac{3su}{t}
\right.\nonumber\\
&&+\left.\frac{2t^3}{su}+\frac{2u^3}{st}+\frac{6tu}{s}+6s+9t+9u\right)+{\cal O}(\epsilon)\,,
\\
E^0_{q,q'\bar{q}'}&=&\frac{1}{(Q^2)^2}\left(\frac{t^2}{s}+\frac{u^2}{s}+t+u\right)+{\cal O}(\epsilon)\,,
\\
E^0_{q,qq'}&=&-\frac{1}{(Q^2)^2}\left(\frac{s^2}{t}+\frac{u^2}{t}+s+u\right)+{\cal O}(\epsilon)\,.
\end{eqnarray}
When integrated over the factorized phase space, they yield
\begin{eqnarray}
{\cal D}^{0}_{q,gg}&=&-4{\bf I}^{(1)}_{qg}(Q^2)\delta(1-x)+(Q^2)^{-\eps}\left[
-\frac{1}{\eps}p^{(0)}_{qq}(x)+\left(\frac{67}{18}-\frac{1}{3}\pi^2\right)\delta(1-x)
-\frac{11}{6}\Dcz(x)
\right.\nonumber\\
&&\left.
+2\Dco(x)
-\frac{1}{3x}+1-x-(1+x)\log(1-x)-\frac{1+x^2}{1-x}\log(x)+{\cal O}(\eps)\right]\,,\\
{\cal E}^{0}_{q,q'\bar{q}'}&=&-4{\bf I}^{(1)}_{qg,F}(Q^2)\delta(1-x)+(Q^2)^{-\eps}\left[
-\frac{5}{9}\delta(1-x)+\frac{1}{3}\Dco(x)-\frac{1}{6x}+{\cal O}(\epsilon)\right]\,,\\
{\cal E}^{0}_{q,qq'}&=&(Q^2)^{-\eps}\left[-\frac{1}{2\eps}p^{(0)}_{gq}(x)+
\frac{2}{x}-\frac{3}{2}-\frac{(2-2x+x^2)}{2x}\log\left(\frac{1-x}{x}\right)+{\cal O}(\epsilon)\right]\,.
\end{eqnarray}
Finally, there is one gluon-gluon antenna with a quark in the initial state:
\begin{equation}
G^0_{q,qg}=-\frac{1}{(Q^2)^2}\left(\frac{s^2}{t}+\frac{u^2}{t}\right)+{\cal O}(\epsilon)\,,
\end{equation}
yielding
\begin{eqnarray}
{\cal G}^{0}_{q,qg}&=&(Q^2)^{-\eps}\left[-\frac{1}{2\eps}p^{(0)}_{gq}(x)-
\frac{7}{4x}+1+\frac{(2-2x+x^2)}{2x}\log\left(\frac{1-x}{x}\right)+{\cal O}(\epsilon)\right]\,,
\end{eqnarray}
when integrated over the antenna phase space.

\subsubsection{Gluon initiated antennae}
\label{sec:gini}
For the gluon initiated antennae, we find one quark-quark antenna 
\begin{equation}
A^{0}_{g,q\bar{q}}=\frac{1}{Q^2}\left(\frac{2s}{t}+\frac{2s}{u}+\frac{2s^2}{tu}+
        \frac{u}{t}+\frac{t}{u}\right)+{\cal O}(\epsilon)\,.
\end{equation}
Its integrated form is
\begin{eqnarray} 
{\cal A}^{0}_{g,q\bar{q}}&=&(Q^2)^{-\eps}\left[-\frac{1}{\eps}p^{(0)}_{qg}(x)-
(1-2x+2x^2)(\log(1-x)-\log(x))+{\cal O}(\epsilon)\right]\,.
\end{eqnarray}
There is one quark-gluon antenna with a gluon in the initial state
\begin{eqnarray}\label{eq:Dqgq-unsplit}
D^{0}_{g,qg}&=&\frac{1}{(Q^2)^2}\left(\frac{u^2}{t}+\frac{u^2}{s}+\frac{4t^2}{s}+\frac{4t^2}{u}+
\frac{4s^2}{u}+\frac{4s^2}{t}+\frac{3tu}{s}+\frac{3su}{t}
\right.\nonumber\\
&+&\left.\frac{2t^3}{su}+\frac{2s^3}{tu}+\frac{6st}{u}+6u+9t+9s\right)+{\cal O}(\epsilon)\,.
\end{eqnarray}
This antenna contains singular limits when the quark or the gluon in the final state
become collinear with the initial state gluon. 
In the first case it collapses into
a quark-gluon antenna and in the second case into a gluon-quark one.
Accordingly, the reduced matrix elements
accompanying these two singular configurations are different. Thus,
the antenna must be 
split to separate these two configurations.
This can be easily done by partial fractioning in the variables $t$ and $u$, we obtain
\begin{eqnarray}
D^{0}_{g,qg}&=&-\frac{1}{2}\frac{1}{(Q^2)^2}\left(\frac{2u^2}{t}+\frac{8s^2}{t}+\frac{6su}{t}
-\frac{4s^3}{t(Q^2+s)}\right)+{\cal O}(\epsilon)\,,
\end{eqnarray}
and
\begin{eqnarray}
D^{0}_{g,gq}&=&\frac{1}{2}\frac{1}{(Q^2)^2}\left(\frac{2t^2}{s}+\frac{8u^2}{s}+\frac{8u^2}{t}+
\frac{8s^2}{t}+\frac{6tu}{s}+\frac{4u^3}{st}-\frac{4s^3}{(Q^2+s)t}
\right.\nonumber\\
&&+\left.\frac{12su}{t}+12t+18u+18s\right)+{\cal O}(\epsilon)\,,
\end{eqnarray}
where we have adjusted the names of the antennae so that $D^{0}_{g,ij}$ now does not contain 
singularities when $j$ becomes collinear with the initial state gluon. We also changed the
sign of the first sub-antenna and exchanged
$t$ and $u$ in the second case to agree with the definitions given at the beginning of the 
section. The two sub-antennae can be integrated over the factorized phase space, namely:
\begin{eqnarray}
{\cal D}^{0}_{g,qg}&=&(Q^2)^{-\eps}\left[
-\frac{1}{2\,\eps}p^{(0)}_{qg}(x)+\frac{3}{4x}-1+\frac{1}{2}(1-2x+2x^2)\log(1-x)\right.\nonumber\\
&&-\frac{1}{2}\left.(1-2x+2x^2)\log(x)+
{\cal O}(\eps)\right]\,.
\end{eqnarray}
and
\begin{eqnarray}
{\cal D}^{0}_{g,gq}&=&-2{\bf I}^{(1)}_{qg}(Q^2)\delta(1-x)+(Q^2)^{-\eps}\left[
-\frac{1}{2\eps}p^{(0)}_{gg}(x)+\left(\frac{7}{4}-\frac{1}{6}\pi^2\right)\delta(1-x)\right.\nonumber\\
&&-\left.\frac{3}{4}\Dcz(x)+\Dco(x)-\frac{3}{2}+\frac{1-2x+x^2-x^3}{x}\log(1-x)\right.\nonumber\\
&&-\left.\frac{(1-x+x^2)^2}{x(1-x)}\log(x)+
{\cal O}(\eps)\right]\,.
\end{eqnarray}
Finally there are two gluon-gluon antennae
\begin{eqnarray}
F^{0}_{g,gg}&=&\frac{1}{2}\frac{1}{(Q^2)^2}\left(\frac{8s^2}{t}+\frac{8t^2}{s}+\frac{8s^2}{u}+\frac{8u^2}{s}
+\frac{8t^2}{u}+\frac{8u^2}{t}+\frac{12st}{u}+\frac{12su}{t}+\frac{12tu}{s}\right.\nonumber\\
&&+\left.\frac{4t^3}{su}+\frac{4u^3}{st}+\frac{4s^3}{tu}+24s+24t+24u\right)+{\cal O}(\epsilon)\,,\\
G^0_{g,q\bar{q}}&=&\frac{1}{(Q^2)^2}\left(\frac{t^2}{s}+\frac{u^2}{s}\right)+{\cal O}(\epsilon)\,.
\end{eqnarray}
Their integrated forms are given by:
\begin{eqnarray}
{\cal F}^{0}_{g,gg}&=&-4{\bf I}^{(1)}_{gg}(Q^2)\delta(1-x)+(Q^2)^{-\eps}\left[
-\frac{1}{\eps}p^{(0)}_{gg}(x)+\left(\frac{67}{18}-\frac{1}{3}\pi^2\right)\delta(1-x)
-\frac{11}{6}\Dcz(x)
\right.\nonumber\\
&&\left.
+2\Dco(x)
-\frac{11}{6x}+\frac{2(1-2x+x^2-x^3)}{x}\log(1-x)-\frac{2(1-x+x^2)^2}{x(1-x)}\log(x)
\right.\nonumber\\
&&\left.
+{\cal O}(\eps)\right.\bigg]\,,\\
{\cal G}^{0}_{g,q\bar{q}}&=&-2{\bf I}^{(1)}_{gg,F}(Q^2)\delta(1-x)+(Q^2)^{-\eps}\left[
-\frac{5}{9}\delta(1-x)+\frac{1}{3}\Dco(x)+\frac{1}{3x}+{\cal O}(\epsilon)\right]\,.
\end{eqnarray}
\section{Initial-initial configurations}
\label{sec:ii}

The last situation to be considered is when the two hard radiators are in the
initial state. The subtraction terms necessary to account for singularities 
associated with these configurations are constructed in terms of
initial-initial antennae. At NLO, one unresolved parton is emitted off these 
two radiators, as  displayed in Figure~\ref{fig:ii}. As before, 
more final state partons can be emitted at higher orders. 

The initial-initial configuration is slightly more involved than the
previous two. Even though at NLO the integration of the antenna functions
over the factorized phase space will turn out to be trivial, in order to
guarantee this factorization, a very restricted kind of mappings will be
allowed. In addition, to fulfill overall momentum conservation, both the
hard radiators and {\em all} the spectator momenta, including non-colored
particles, have to be remapped. This is done with a convenient generalization
of the Lorentz transformation introduced in \cite{Catani:1996vz}.

\subsection{Subtraction terms for initial-initial configurations}

The NLO antenna subtraction term, to be convoluted with the
appropriate parton distribution functions for the initial state
partons,  for a configuration with the two hard emitters in the initial state 
(partons $i$ and $k$ with momenta $p_1$ and 
$p_2$) can be 
written as: 
\begin{eqnarray}\label{eq:subii}
d\hat\sigma^{S,(ii)}&=&{\cal N}\sum_{m+1}\d\Phi_{m+1}(k_1,\dots,k_{j-1},k_j,k_{j+1},\dots,k_{m+1};p_1,p_2)
  \,\frac{1}{S_{m+1}}\nonumber \\
&& \sum_{j}X^{0}_{ik,j}(p_1,p_2,k_j)
  \left|{\cal M}_m(\kt_1,\dots,\kt_{j-1},\kt_{j+1},\dots,\kt_{m+1}; x_1p_1,x_2p_2)\right|^2\nonumber\\
&&\times  J^{(m)}_{m}(\kt_1,\dots,\kt_{j-1},\kt_{j+1},\dots,\kt_{m+1})\,.  
\end{eqnarray}
As mentioned, all the momenta in the arguments of the reduced
matrix elements and the jet functions have been redefined.
The two hard radiators are simply rescaled by factors 
$x_1$ and $x_2$ respectively. The spectator momenta are boosted 
by a Lorentz transformation onto the new set of 
momenta $\{\kt_l,\,l\neq j\}$. As before, the mapping 
must satisfy overall momentum conservation and keep the mapped momenta 
in the mass shell. In this case, this turns out to severely restrict the
possible mappings.

We start from the $(m+1)$-parton phase space 
\begin{equation}
\d\Phi_{m+1}(k_1,\dots,k_{m+1};p_1,p_2)=(2\pi)^{d}\delta\left(p_1+p_2-\sum_l k_l\right)
\,\prod_{l}[\d k_l]\,
\end{equation}
and insert
\begin{equation}
1=\int \d^d q  \d^d \tilde{q}\,\delta\left(p_1+p_2-k_j-q\right)\,
\delta\left(x_1p_1+x_2p_2-\tilde{q}\right)\,,
\end{equation}
and
\begin{equation}
1=\int \prod_{l\neq j}\delta(\kt_l-B(k_l,q,\tilde{q}))[\d\kt_l]\,,
\end{equation}
where $B$ is a Lorentz transformation that maps $q$ onto
$\tilde{q}$. We also insert
\begin{equation}
1=\int \d x_1 \d x_2\delta(x_1-\hat{x}_1)\delta(x_2-\hat{x}_2)\,
\end{equation}
with
\begin{eqnarray}\label{eq:mapii}
\hat{x}_1&=&\left(\frac{s_{12}-s_{j2}}{s_{12}}\,\frac{s_{12}-s_{1j}-s_{j2}}{s_{12}-s_{1j}}\right)^{\frac{1}{2}}\,,\nonumber \\
\hat{x}_2&=&\left(\frac{s_{12}-s_{1j}}{s_{12}}\,\frac{s_{12}-s_{1j}-s_{j2}}{s_{12}-s_{j2}}\right)^{\frac{1}{2}}\,.
\end{eqnarray}
These last two definitions guarantee the overall momentum conservation in the
mapped momenta and the right soft and collinear behavior, they are derived in detail in 
Section~\ref{sec:mapii} below. 
Now we can integrate over the original momenta, $k_l\,,l\neq j$ by inverting
the Lorentz transformation. The Jacobian factor 
associated with this integration is unity, as $B$ is a proper Lorentz 
transformation. We also integrate over the auxiliary momenta $q$ and
$\tilde{q}$,
to obtain
\begin{eqnarray}
\d\Phi_{m+1}(k_1,\dots,k_{m+1};p_1,p_2)&=&
\d\Phi_{m}(\kt_1,\dots,\kt_{j-1},\kt_{j+1},\dots,\kt_{m+1};x_1p_1,x_2p_2)
\nonumber\\
&&\times\delta(x_1-\hat{x}_1)\,\delta(x_2-\hat{x}_2)\,[\d k_j]\,\d x_1\,\d x_2\,.
\end{eqnarray}
At this point the phase space totally factorized into the convolution of 
an $m$ particle phase space, involving only the redefined momenta, 
with the phase space of parton $j$. 

Inserting the factorized expression for the phase space measure in eq.
(\ref{eq:subii}), the subtraction terms can be integrated over
the antenna phase space. The integrated form of the
subtraction terms must be, then, combined with the virtual
and mass factorization terms to cancel the explicit poles in $\epsilon$.
In the case of initial-initial subtraction terms, the antenna 
phase space is trivial: the two remaining Dirac delta functions
can be combined with the one particle phase space, such that there
are no integrals left. We define the initial-initial
integrated antenna functions as follows:
\begin{equation}
{\cal X}_{ik,j}(x_1,x_2)=\frac{1}{C(\epsilon)}\int
[\d k_j]\,x_1\,x_2\,\delta(x_1-\hat{x}_1)\,\delta(x_2-\hat{x}_2)\,X_{ik,j}
\end{equation}
Substituting the one particle phase space, and carrying out the
integrations over the Dirac delta functions, we have, 
\begin{eqnarray}
{\cal X}_{ik,j}(x_1,x_2)&=&({Q^2})^{-\epsilon}\frac{e^{\epsilon\gamma_E}}{\Gamma(1-\epsilon)}\,
{\cal J}(x_1,x_2)\,Q^2\,X_{ik,j}\,,
\end{eqnarray}
with $Q^2=q^2=(p_1+p_2-k_j)^2$. The Jacobian factor, ${\cal J}(x_1,x_2)$ is given by
\begin{equation}
{\cal
  J}(x_1,x_2)=\frac{x_1\,x_2\,(1+x_1\,x_2)}{(x_1+x_2)^2}\,(1-x_1)^{-\epsilon}(1-x_2)^{-\epsilon}\,
\left(\frac{(1+x_1)(1+x_2)}{(x_1+x_2)^2}\right)^{-\epsilon}\,,
\end{equation}
and the two-particle invariants are given by:
\begin{eqnarray}
s_{1j}=-s_{12}\frac{x_1\,(1-x_2^2)}{x_1+x_2}\,,\qquad
s_{j2}=-s_{12}\frac{x_2\,(1-x_1^2)}{x_1+x_2}\,.
\end{eqnarray} 
The integrated subtraction term is, then,
\begin{eqnarray}
\d\hat{\sigma}^{S,(ii)}&=&\sum_{m+1}\sum_{j}
\frac{{\cal N}}{S_{m+1}}\int\frac{\d x_1}{x_1}\frac{\d x_2}{x_2}\,
C(\epsilon)\,{\cal X}_{ik,j}(x_1,x_2)
\nonumber\\&&
\times \d\Phi_{m}(k_1,\dots,k_{j-1},k_{j+1},\dots,k_{m+1};x_1p_1,x_2p_2)
\nonumber\\&&
\times\left|{\cal
    M}_m(k_1,\dots,k_{j_1},k_{j+1},\dots,k_{m+1};x_1p1,x_2p_2)\right|^2
\nonumber\\&&
\times J^{(m)}_{m}(k_1,\dots,k_{j-1},k_{j+1},\dots,k_{m+1})\,,
\end{eqnarray}
where we have relabeled all $\tilde{k}_i\to k_i$. 
The final step is to convolute this subtraction term with the parton
distribution functions of the initial state particles. The integrated version
of the subtraction pieces is then combined with the virtual and mass
factorization terms to render a finite contribution when 
$\epsilon\rightarrow 0$. Recasting appropriately the convolutions, the
integrated subtraction term is
\begin{eqnarray}
d{\sigma}^{S,(ii)}&=&\sum_{m+1}\sum_{j}
\frac{S_{m}}{S_{m+1}}\int\frac{\d\xi_1}{\xi_1}\int\frac{\d\xi_2}{\xi_2}
\int_{\xi_1}^{1}\frac{\d x_1}{x_1}
\,\int_{\xi_2}^{1}\frac{\d x_2}{x_2}
f_{i/1}\left(\frac{\xi_1}{x_2}\right)\,f_{k/2}\left(\frac{\xi_2}{x_2}\right)\nonumber\\
&&\times C(\epsilon)\,{\cal X}_{ik,j}(x_1,x_2)\,\d\hat{\sigma}^B(\xi_1 H_1,\xi_2 H_2)\,.
\end{eqnarray}

\subsection{Phase space mapping}
\label{sec:mapii}
By asking for momentum conservation and phase space factorization,
we are severely constraining the possible phase space mapping. The principal 
origin of this constraint is that the remapping of 
both initial state momenta can only be a rescaling, since 
 any transversal component would spoil the
phase space factorization.

The two mapped  initial state momenta must be of the form 
\begin{equation}
P_1=x_1p_1\,\qquad P_2=x_2p_2\, ,
\end{equation} 
so that 
\[\tilde q\equiv P_1+P_2\]
is in the beam axis. Since the vector component of 
$q\equiv p_1+p_2-k_j$ is in general not along the $p_1-p_2$ axis 
we need to boost all the other momenta in order to restore momentum 
conservation.
The transformation must map $q$ onto $\tilde{q}$. As it must 
keep all the spectator momenta, which are 
arbitrary vectors, on mass-shell, it must 
belong to the Lorentz group. This transformation then fully determines 
the initial-initial phase space mapping, by fixing $x_{1,2}$ in terms 
of the invariants.

We consider a candidate Lorenz transformation $\Lambda(q)$. It has 
to map the vector $q$ into a vector $\Lambda(q) q=\tilde q$ in the beam axis.
From the result $\tilde q$ of the transformation, 
one can read off $x_1$ and $x_2$ using
\begin{eqnarray*}
2 (p_1+p_2)\tilde q&=&(x_1+x_2) s_{12}\nonumber\\
2 (p_1-p_2)\tilde q&=&(x_2-x_1) s_{12}
\end{eqnarray*}    
yielding
\begin{eqnarray}
x_1&=&\frac{2 p_2\tilde q}{s_{12}}\nonumber\\
x_2&=&\frac{2 p_1\tilde q}{s_{12}}
\end{eqnarray}
The two equations can be combined to give 
\[x_1 x_2=\frac{s_{12}-s_{1j}-s_{2j}}{s_{12}}\;,\]
which can also be derived from the on shell condition $q^2 = \tilde{q}^2$. 
To ensure that the mapping has the right soft and collinear 
limits at NLO it is sufficient to impose 
$\Lambda(q) =1$ for $q$ in the beam axis.

For the transformation $\Lambda$ we take a boost $B_T^*(q)$ of 
appropriate parameter whose direction is transverse to the beam axis 
 in the rest frame of $P_1$ and $P_2$. Objects defined in 
the rest frame of the new system are denoted by a $^*$. This transformation 
clearly satisfies the requirement
$B_T^*(q) =1$  for $q$ in the beam axis, 
since then no boost is required to bring $q$ in the beam axis. 
By construction, the longitudinal component of 
$q$ in the rest frame of $P_1$ and $P_2$ is conserved, that is
\begin{equation}
q_L^*=\frac{(P_1-P_2)q}{x_1 x_2\sqrt{s}}=\tilde q_L^*=\frac{(P_1-P_2)\tilde q}{x_1 x_2\sqrt{s}}\;.
\end{equation} 
So that we have 
\begin{equation}
(x_2-x_1)=\frac{x_2 s_{2j}-x_1 s_{1j}}{s_{12}}
\end{equation}
which gives the mapping
\begin{eqnarray}
x_1 =  \sqrt{\frac{s_{12}-s_{2j}}{s_{12}-s_{1j}}}\sqrt{\frac{s_{12}-s_{1j}-s_{2j}}{s_{12}}},\nonumber\\
x_2 = \sqrt{\frac{s_{12}-s_{1j}}{s_{12}-s_{2j}}}\sqrt{\frac{s_{12}-s_{1j}-s_{2j}}{s_{12}}} \;,
\end{eqnarray}
which was used in (\ref{eq:mapii}) above. It yields the correct soft and 
collinear limits at NLO:
\begin{enumerate}
\item $j$ soft: $x_1 \to 1$, $x_2 \to 1$.
\item $k_{j} = z_1p_1 \parallel p_1$: $x_1 = (1-z_1)$, $x_2 = 1$.  
\item $k_{j} = z_2p_2 \parallel p_2$: $x_1 = 1$, $x_2 = (1-z_2)$.  
\end{enumerate}

It should be pointed out the transformation is not unique. Possible
transformations are however strongly constrained.
If one requires a symmetrical treatment of $x_1$ and $x_2$, 
rotations are not allowed as transformation. To show that, 
we take $p_j$ to be transverse to the beam axis. Bringing $q$ to the 
beam axis with a rotation will force us to choose to rotate $\vec q$ either 
towards the $p_1$ or the $p_2$ side. This would favor either $x_1$ or $x_2$. 
The only way to bring $q$ to the beam axis, without 
having to choose between $x_1$ and $x_2$ is in this case a boost 
transverse to the beam axis.

The extension of the phase space mapping to NNLO is trivial. In this case,
four-parton antenna functions $X_{il,jk}$ require a mapping with two partons, 
$j$ and $k$ unresolved. The transformation $B_T^*(q)$ is unchanged, but now the vector $q$ is given
by $q=p_1+p_2-k_{j}-k_{k}$. The momentum fractions in 
(\ref{eq:mapii}) are replaced by
\begin{eqnarray}
x_1&=&\left(\frac{s_{12}-s_{j2}-s_{k2}}{s_{12}}
\;    \frac{s_{12}-s_{1j}-s_{1k}-s_{j2}-s_{k2}+s_{jk}}{s_{12}-s_{1j}-s_{1k}}
     \right)^{\frac{1}{2}}\,,\nonumber\\
x_2&=&\left(\frac{s_{12}-s_{1j}-s_{1k}}{s_{12}}
\;    \frac{s_{12}-s_{1j}-s_{1k}-s_{j2}-s_{k2}+s_{jk}}{s_{12}-s_{j2}-s_{k2}}
     \right)^{\frac{1}{2}}\,.
\end{eqnarray}
These two momentum fractions satisfy the following limits in 
double unresolved 
configurations:
\begin{enumerate}
\item $j$ and $k$ soft: $x_1\rightarrow 1$, $x_2\rightarrow 1$,
\item $j$ soft and $k_{k}=z_1p_1\parallel p_1$: $x_1\rightarrow 1-z_1$, $x_2\rightarrow 1$,
\item $k_{j}=z_1p_1\parallel p_1$ and $k_{k}=z_2p_2\parallel p_2$:
  $x_1\rightarrow 1-z_1$, $x_2\rightarrow 1-z_2$,
\item $k_{j}+k_{k}=z_1p_1\parallel p_1$: $x_1\rightarrow 1-z_1$,
  $x_2\rightarrow 1$,
\end{enumerate}
and all the limits obtained from the ones above by exchange of $p_1$ with
$p_2$ and of $k_j$ with $k_k$.
The factorization of the phase space into an $m$-parton phase space and 
an antenna phase space goes along the same lines as for the NLO case. 
At NNLO, however, the integration of the antenna functions over 
this factorized phase space is no longer trivial. 

As in the initial-final case, we also require the NNLO mapping to 
turn into the NLO mapping (\ref{eq:mapii}) if only one parton becomes 
unresolved. In the limits where $j$ becomes unresolved between $i$ and $k$, 
we denote the parameters 
of the reduced NLO phase space mapping by $x_1^\prime$ and $x_2^\prime$. 
We find:
\begin{enumerate}
\item $j$ becomes soft: \begin{eqnarray*}
x_1 &\to& \left(\frac{s_{12}-s_{k2}}{s_{12}}\,\frac{s_{12}-s_{1k}-s_{k2}}{s_{12}-s_{1k}}\right)^{\frac{1}{2}}
 = {x}_1^\prime \,,
\\ x_2 &\to& 
\left(\frac{s_{12}-s_{1k}}{s_{12}}\,\frac{s_{12}-s_{1k}-s_{k2}}{s_{12}-s_{k2}}\right)^{\frac{1}{2}} = {x}_2^\prime \end{eqnarray*}
\item $k_j\parallel k_k$, $k_j+k_k = K_K$: \begin{eqnarray*}
{x}_1 &\to& \left(\frac{s_{12}-s_{K2}}{s_{12}}\,\frac{s_{12}-s_{1K}-s_{K2}}{s_{12}-s_{1K}}\right)^{\frac{1}{2}} = {x}_1^\prime \,, \\
{x}_2 &\to& \left(\frac{s_{12}-s_{1K}}{s_{12}}\,\frac{s_{12}-s_{1K}-s_{K2}}{s_{12}-s_{K2}}\right)^{\frac{1}{2}}= {x}_2^\prime \,. \end{eqnarray*}
\item $k_j = z_1p_1 \parallel p_1$: 
\begin{eqnarray*}
{x}_1 &\to& \left(\frac{(1-z_1) s_{12}-s_{k2}}{s_{12}}\,\frac{(1-z_1)s_{12}
-(1-z_1)s_{1k}-s_{k2}}{s_{12}-s_{1k}}\right)^{\frac{1}{2}} 
= (1-z_1){x}_1^\prime \,, \\
{x}_2 &\to& \left(\frac{s_{12}-s_{1k}}{s_{12}}\,\frac{
(1-z_1)s_{12}-(1-z_1)s_{1k}-s_{k2}}{(1-z_1)s_{12}-s_{k2}}\right)^{\frac{1}{2}}= {x}_2^\prime \,. \end{eqnarray*}
\end{enumerate}
All other single unresolved limits involving 
one radiator parton in the initial state 
follow by exchange of $p_1$ with $p_2$ or 
$k_j$ with $k_k$. To subtract all 
unresolved limits of parton $j$ between emitter partons $i$ and $k$ 
from $X_{il,jk}$, one needs to subtract from it the 
product of an initial-final antenna function with an initial-initial 
antenna function $X_{i,jk}\cdot X_{Il,K}$. Analytic integration of these terms 
over both antenna phase spaces results in a double convolution in 
the rescaling variables for $p_1$ and a single convolution in the 
rescaling variable for $p_2$. 

At subleading colour, $j$ can also become unresolved between $i$ and $l$. 
In this case, we denote the reduced phase space mapping parameters by
$x_1^{\prime\prime}$ and $x_2^{\prime\prime}$. The limits read:
\begin{enumerate}
\item $j$ becomes soft: \begin{eqnarray*}
x_1 &\to& \left(\frac{s_{12}-s_{k2}}{s_{12}}\,\frac{s_{12}-s_{1k}-s_{k2}}{s_{12}-s_{1k}}\right)^{\frac{1}{2}}
 = {x}_1^{\prime\prime} \,,\\
 x_2 &\to& 
\left(\frac{s_{12}-s_{1k}}{s_{12}}\,\frac{s_{12}-s_{1k}-s_{k2}}{s_{12}-s_{k2}}\right)^{\frac{1}{2}} = {x}_2^{\prime\prime} \end{eqnarray*}
\item $k_j = z_1p_1 \parallel p_1$: 
\begin{eqnarray*}
{x}_1 &\to& \left(\frac{(1-z_1) s_{12}-s_{k2}}{s_{12}}\,\frac{(1-z_1)s_{12}
-(1-z_1)s_{1k}-s_{k2}}{s_{12}-s_{1k}}\right)^{\frac{1}{2}} 
= (1-z_1){x}_1^{\prime\prime}  \,, \\
{x}_2 &\to& \left(\frac{s_{12}-s_{1k}}{s_{12}}\,\frac{
(1-z_1)s_{12}-(1-z_1)s_{1k}-s_{k2}}{(1-z_1)s_{12}-s_{k2}}\right)^{\frac{1}{2}}= {x}_2^{\prime\prime} \,. \end{eqnarray*}
\item $k_j = z_2p_2 \parallel p_2$: 
\begin{eqnarray*}
{x}_1 &\to& \left(\frac{s_{12}-s_{k2}}{s_{12}}\,\frac{(1-z_2)s_{12}
-s_{1k}-(1-z_2)s_{k2}}{(1-z_2)s_{12}-s_{1k}}\right)^{\frac{1}{2}} 
= {x}_1^{\prime\prime}  \,, \\
{x}_2 &\to& \left(\frac{(1-z_2)s_{12}-s_{1k}}{s_{12}}\,\frac{
(1-z_2)s_{12}-s_{1k}-(1-z_2)s_{k2}}{s_{12}-s_{k2}}\right)^{\frac{1}{2}}= (1-z_2) {x}_2^{\prime\prime}  \,. \end{eqnarray*}
\end{enumerate}
These single unresolved limits are subtracted from 
$X_{il,jk}$ by the product of two initial-initial antenna functions 
$X_{il,j}\cdot X_{IL,k}$. 
Analytic integration of these terms 
over both antenna phase spaces results in two  double convolutions in 
the rescaling variables for $p_1$ and  $p_2$.

\subsection{NLO antenna functions}\label{sec:antenna-ii}
The unintegrated antenna functions necessary to subtract all the 
singular configurations at NLO with two initial state hard radiators, 
can be obtained immediately from the corresponding initial-final
ones quoted in section \ref{sec:antenna-if} by crossing. We have
\begin{equation}
X^0_{ik,j}=\delta_{ik,j}\,X^0_{j,ik}\,,
\end{equation}
where $\delta_{ik,j}$ is an overall sign, 
which is $-1$ for  $D^0_{gg,q}$ and $E^0_{q\bar{q},q'}$ and $+1$ for 
 all other antennae. 

The Mandelstam variables of 
the unintegrated antennae in section \ref{sec:antenna-if} have to be 
replaced by $s=(p_i+p_k)^2$, $t=(p_i-p_j)^2$, $u=(p_k-p_j)^2$.

Again, the splitting of antenna functions into different sub-antennae 
is different from the two configurations discussed above.
For the initial-initial configurations there is no need to split the
quark-gluon antenna $D^0_{qg,g}$ as in all the singular limits it collapses
to the same two-particle antenna. 
So $D^0_{qg,g}$ is given by the crossing 
of (\ref{eq:Dqgq-unsplit}). However, in this case, the antenna $D^0_{gg,q}$ 
has to be split into two subantennae to separate the two collinear 
limits present in it. We find:
\begin{equation}
D^{0}_{gg,q}=D^{0}_{g_1g_2,q}+D^{0}_{g_2g_1,q}\,,
\end{equation} 
such that $D^{0}_{g_1g_2,q}$ contains only the singular configurations when
the quark becomes collinear with gluon $g_2$. Explicitly:
\begin{equation}
D^{0}_{g_1g_2,q}=\frac{1}{(Q^2)^2}\left(\frac{s^2}{u}+\frac{4t^2}{u}+
\frac{4u^2}{s}+\frac{3st}{u}
+\frac{2t^3}{su}+\frac{3tu}{s}+3s+9u\right)+{\cal O}(\epsilon)\,,
\end{equation}
As mentioned, the integration of the initial-initial antennae over the 
factorized phase space is trivial and only involves a proper treatment of
the singularities when $\epsilon\rightarrow 0$. The integrated antennae 
read
\begin{eqnarray}
{\cal A}^{0}_{qg,q}&=&(Q^2)^{-\epsilon}\bigg[
-\frac{1}{2\epsilon}p^{(0)}_{qg}(x_2)\,\delta(1-x_1)\nonumber\\
&&+\left(\frac{1}{2}+\frac{1-2 x_2+2 x_2^2}{2} \log \left(2\frac{
   1-x_2}{1+x_2}\right)\right) \delta (1-x_1)\nonumber\\
&+&\Dcz(x_1) \frac{1-2x_2 + 2 x_2^2}{2}\nonumber\\
&-&\frac{1-2 x_2+2 x_2^2}{2(1-x_1)}+\frac{x_1 (1+x_1 x_2) \left(2
   x_1 x_2^3+x_2^2+\left(2 x_2^4-2 x_2^2+1\right) x_1^2\right)}{(x_1+x_2)^3 \left(1-x_1^2\right)}\nonumber\\
   &+&\frac{
   x_1 x_2 (1+x_1 x_2) \left(2 x_1-x_2 \left(x_1^2+2 x_2
   x_1-1\right)\right)}{(x_1+x_2)^3}
\bigg]\,,
\\
{\cal
  A}^{0}_{q\bar{q},g}
&=&-{\bf I}^{(1)}_{q\bar q}(Q^2)\delta(1-x_1)\delta(1-x_2)
+(Q^2)^{-\epsilon}\bigg[
-\frac{1}{2\epsilon}p^{(0)}_{qq}(x_1)\delta(1-x_2)\bigg]
\nonumber\\
&&+(Q^2)^{-\epsilon}\bigg[ 
\frac{\left(x_1^2+x_2^2\right) \left(x_2^2 x_1^2+2 x_2 x_1^2+x_1^2+3 x_2 x_1+2 x_1+1\right)}{2(x_1+1) (x_2+1) (x_1+x_2)^2}\nonumber\\
&&+\frac{\left((1-x_1)^2-2 x_1^2 \log
   \left(\frac{1+x_1}{2}\right)+\left(1-x_1^2\right) \log \left(\frac{2}{1-x_1^2}\right)\right) \delta (1-x_2)}{2(1-x_1)}\nonumber\\
 &&+\frac{1}{4} \pi ^2 \delta (1-x_1) \delta (1-x_2)-\frac{1}{2}(1+x_2) \Dcz(x_1)+\frac{1}{2}\Dcz(x_1) \Dcz(x_2)
\nonumber\\
&&+\delta (1-x_1) \Dco(x_2)
+(x_1\leftrightarrow x_2)\bigg]\,,
\\
{\cal D}^{0}_{qg,g}
&=&-2\left({\bf I}^{(1)}_{qg}(Q^2)\right)\delta(1-x_1)\delta(1-x_2)\nonumber\\
&&+(Q^2)^{-\epsilon}\bigg[
-\frac{1}{2\epsilon}p^{(0)}_{qq}(x_1)\delta(1-x_2)-\frac{1}{2\epsilon}p^{(0)}_{gg}(x_2)\delta(1-x_1)\nonumber\\
&&+\frac{\pi^2}{2}
   \delta (1-x_1) \delta (1-x_2)+ \delta (1-x_2) \Dco(x_1)+ \delta (1-x_1) \Dco(x_2)\nonumber\\
&&+\left(\frac{1-x_1}{2}+\frac{\log
   \left(\frac{2}{x_1+1}\right)}{1-x_1}-\frac{1+x_1}{2} \log \left(2\frac{ 1-x_1}{x_1+1}\right)\right) \delta (1-x_2)\nonumber\\
&&+ \left(-x_2^2+x_2-2+\frac{1}{x_2}\right)  \Dcz(x_1)-\frac{1+x_1}{2} \Dcz(x_2)+
   \Dcz(x_1) \Dcz(x_2)\nonumber\\
&&+\frac{ (x_1 x_2-1) \left(x_2^2 x_1^4+\left(2 x_2^3+x_2\right) x_1^3+\left(2 x_2^4-x_2^2+4\right)
   x_1^2\right) (x_1 x_2+1)^2}{x_1 \left(1-x_1^2\right) (x_1+x_2)^3\left(1-x_2^2\right)}\nonumber\\
&&+\frac{ (x_1 x_2-1) \left(5 x_1 x_2+2 x_2^2\right) (x_1 x_2+1)^2}{x_1 \left(1-x_1^2\right) (x_1+x_2)^3\left(1-x_2^2\right)}\nonumber\\
&&+\frac{ x_2^2- x_2+2-\frac{1}{x_2}}{1-x_1}+\frac{4-x_1 x_2 \left(x_2+x_1
   \left(x_1^2+x_1+1\right) (x_2+1)-3\right)}{2x_1 x_2 \left(1-x_1^2\right) \left(1-x_2^2\right)}\nonumber\\
&&+\left(\frac{ \log
   \left(\frac{2}{x_2+1}\right)}{1-x_2}-\frac{ \left(x_2^3-x_2^2+2 x_2-1\right) \log \left(2\frac{
   1-x_2}{x_2+1}\right)}{x_2}\right) \delta (1-x_1)
\bigg]\,,\\
{\cal D}^{0}_{g_1g_2,q}&=&(Q^2)^{-\epsilon}\bigg[-\frac{1}{2\epsilon}p^{(0)}_{qg}(x_2)\delta(1-x_1)
\nonumber\\
&&+(Q^2)^{-\epsilon}\left(\frac{1}{2}+\frac{1-2 x_2+2 x_2^2}{2} \log \left(2\frac{1-x_2}{1+x_2}\right)\right) \delta
   (1-x_1)\nonumber\\
&&+\frac{ \left(2 x_2 x_1^3+x_1^2+x_2^2 \left(2 x_1^4-2 x_1^2+1\right)\right) (1+x_2 x_1)^2}{\left(1-x_2^2\right) (x_2+x_1)^4}\nonumber\\
&&-\frac{\left(4 x_1^2 x_2^4+3 \left(2 x_1^3+x_1\right)
   x_2^3+\left(4 x_1^4+2 x_1^2+1\right) x_2^2+3 x_1^3 x_2+x_1^2\right) (x_2
   x_1+1)}{2(x_2+x_1)^4}\nonumber\\
&&+\frac{\left(2 x_1^4+2 x_1^3-x_1^2+1\right) \Dcz(x_2)}{2(1+x_1)^2}-\frac{1-2 x_2+2 x_2^2}{2(1-x_1)}\bigg]\,,
\\
{\cal
  E}^{0}_{q\bar{q},q'}&=&(Q^2)^{-\epsilon}\frac{x_1\,x_2\,(1+x_1\,x_2)^2\,(1-x_1^2)}
{2(x_1+x_2)^4}
+(x_1\leftrightarrow x_2)\,,\\
{\cal E}_{qq',q}
&=&(Q^2)^{-\epsilon}\bigg[
-\frac{1}{2\epsilon}p^{(0)}_{gq}(x_1)\delta(1-x_2)\nonumber\\
&&+\frac{\Dcz(x_2) \left(2-2 x_1+x_1^2\right)}{2 x_1}-\frac{x_1^2-2 x_1+2}{2 x_1 (1-x_2)}\nonumber\\
&& -\frac{x_2 (x_1
   x_2+1) \left((x_1 x_2-1) x_1^2+2\right)}{x_1 (x_1+x_2)^2 \left(x_2^2-1\right)}\nonumber\\
   &&+\frac{\left(2
   x_1+\left(2-2 x_1+x_1^2\right) \log \left(\frac{2 (1-x_1)}{x_1+1}\right)-2\right) \delta (1-x_2)}{2 x_1}
\bigg]\,,
\\
{\cal F}^{0}_{gg,g}&=&-{\bf I}^{(1)}_{gg}(Q^2)\delta(1-x_1)\delta(1-x_2)+
(Q^2)^{-\epsilon}\bigg[-\frac{1}{2\epsilon}p^{(0)}_{gg}(x_1)\delta(1-x_2)\bigg]
\nonumber\\&&+\frac{1}{4} \pi ^2 \delta (1-x_1) \delta (1-x_2)+ \left(-x_2^2+x_2-2+\frac{1}{x_2}\right)
   \Dcz(x_1)+\delta (1-x_2) \Dco(x_1)\nonumber\\
   &&+\left(\frac{ 1+\left(\log \left(\frac{2}{x_2+1}\right)-1\right)}{1-x_2}-\frac{
   \left(x_2^3-x_2^2+2 x_2-1\right) \left(\log \left(2\frac{ 1-x_2}{x_2+1}\right)\right)}{x_2}\right) \delta
   (1-x_1)\nonumber\\
&&+\frac{1}{2}\Dcz(x_1) \Dcz(x_2)-\frac{3}{2(1+x_1) (1+x_2)}+
\frac{2 \left(x_1^4-x_1^2+2\right) (1+x_1 x_2)^2}{(x_1+x_2)^4}+\nonumber\\
&&+\frac{ x_1^2- x_1+2+\frac{2}{(x_2+1) x_1}-\frac{1}{x_1}}{1-x_2}-\frac{2 \left(x_1^4+x_1^2+\left(x_1^2+3\right) x_1+2\right) (1+x_1 x_2)^2}{(1-x_2^2)
   (x_1+x_2)^4}\nonumber\\
&&+\frac{3 \left(x_1^3+x_1\right) \left(x_2^3+x_2\right) (1+x_1 x_2)}{(x_1+x_2)^4}+\frac{2 \left(x_2^2 x_1^4+\left(x_2^2+1\right)^2 x_1^2+x_2^2\right) (1+x_1  x_2)}{(x_1+x_2)^4}\nonumber\\
&&+\frac{2 x_1 \left(x_1^2+3\right) (1+x_1 x_2)^2}{(x_2+1) (x_1+x_2)^4}+\frac{(1+x_1 x_2)}{x_1 (x_1+1) x_2 (x_2+1)}
+(x_1\leftrightarrow x_2)\,,
\\
{\cal G}^{0}_{q\bar{q},g}&=&(Q^2)^{-\epsilon}\frac{(1+x_1\,x_2)x_1^2(1-x_2)^2(1+x_2)^2}{(x_1+x_2)^4}
+(x_1\leftrightarrow x_2)\,,
\\
{\cal G}^{0}_{gq,q}
&=&(Q^2)^{-\epsilon}\bigg[
-\frac{1}{2\epsilon}p^{(0)}_{gq}(x_1)\delta(1-x_2)\nonumber\\
&&-\frac{\left(2 (1-x_1)-\left(2-2 x_1+x_1^2\right) \log \left(2\frac{1-x_1}{x_1+1}\right)\right)
   \delta (1-x_2)}{2 x_1}+\frac{\Dcz(x_2) \left(2-2 x_1+x_1^2\right)}{2 x_1}\nonumber\\
&&-\frac{2-2 x_1+x_1^2}{2x_1 (1-x_2)}+\frac{2 (1+x_1 x_2)
   \left(x_1^2+2 x_2 x_1+\left(x_1^4-2 x_1^2+2\right) x_2^2\right)}{2 x_1 (x_1+x_2)^3
   \left(1-x_2^2\right)}
\bigg]\,.
\end{eqnarray}

\section{Application of method at NLO}
\label{sec:appl}
To illustrate the application of the antenna subtraction method at NLO, 
we derive the antenna subtraction terms for two example reactions: deep 
inelastic (2+1)-jet production and hadronic vector-boson-plus-jet production.
Several NLO calculations are already available in the literature
both for the DIS jet 
production~\cite{Catani:1996vz,disjet} 
and for the vector boson production~\cite{vecjet}.

\subsection{(2+1)-jet production  in deep inelastic scattering}
The production of (2+1) jets in deep inelastic lepton-proton scattering 
can be described on the parton level by the scattering of a space-like 
virtual gauge boson and a parton, yielding a final state with two hard partons 
(with the extra jet coming from the proton remnant, not participating in the 
hard interaction). 
 We limit ourselves to consider the scattering of
transversely polarized virtual photons. The cross section can be written as
\begin{equation}
d\sigma=\int\frac{d\xi}{\xi}\sum_q f_q(\xi)\,\d\hat{\sigma}_q+f_g(\xi)\,\d\hat{\sigma}_g\,.
\end{equation}
The partonic cross sections up to NLO are, in turn, given by,
\begin{eqnarray}
\d\hat{\sigma}_q&=&\d\Phi_{2}(k_{g},k_{q};p_q,q)\,\left|M^0_{q,gq}\right|^2\,J^{(2)}_{2}(k_{g},k_{q})\nonumber\\
               &&+\d\Phi_{2}(k_{g},k_{q};p_q,q)\,2\Re(M^{1\dagger}_{q,gq}M^0_{q,gq})\,J^{(2)}_{2}(k_{g},k_{q})\nonumber\\
               &&+\d\Phi_{3}(k_{g_1},k_{g_2},k_{q};p_q,q)\,\left|M^0_{q,ggq}\right|^2\,J^{(3)}_{2}(k_{g_1},k_{g_2},k_{q})\nonumber\\
               &&+\d\Phi_{3}(k_{q_1},k_{q_2},k_{\bar{q}};p_q,q)\,\left|M^0_{q,qq\bar{q}}\right|^2\,J^{(3)}_{2}(k_{q},k_{q},k_{\bar{q}})\nonumber\\
               &&+\sum_{q'\neq
               q}\d\Phi_{3}(k_{q},k_{q'},k_{\bar{q}'};p_q,q)\,\left|M^0_{q,qq'\bar{q}'}\right|^2\,J^{(3)}_{2}(k_{q},k_{q'},k_{\bar{q}'})\,,\\
\d\hat{\sigma}_g&=&\d\Phi_{2}(k_{q},k_{\bar{q}};p_g,q)\,\left|M^0_{g,q\bar{q}}\right|^2\,J^{(2)}_{2}(k_{q},k_{\bar{q}})\nonumber\\
               &&+\d\Phi_{2}(k_{q},k_{\bar{q}};p_q,q)\,2\Re(M^{1\dagger}_{g,q\bar{q}}M^{0}_{g,q\bar{q}})\,J^{(2)}_{2}(k_{q},k_{\bar{q}})\nonumber\\
               &&+\d\Phi_{3}(k_{g},k_{q},k_{\bar{q}};p_q,q)\,\left|M^0_{g,gq\bar{q}}\right|^2\,J^{(3)}_{2}(k_{g},k_{q},k_{\bar{q}})\,. 
\end{eqnarray}
For the sake of brevity, the momentum arguments on the matrix elements 
were omitted. The first index on the matrix elements indicates the 
incoming parton, the remaining ones the outgoing partons.
Subtractions for infrared real radiation singularities 
must be performed only on the three-parton final states. 
The matrix elements for the real contributions can be expressed in terms 
of colour-ordered three-parton and four-parton antenna functions. They read 
as follows:
\begin{eqnarray}
\left|M^0_{q,gq}\right|^2&=&e_q^2\,N_{2,q}\,A_3^0(1_q,3_g,\hat{2}_{\bar{q}})\\
\left|M^0_{q,ggq}\right|^2&=&e_q^2\,\frac{1}{2}\,N_{3,q}\,\bigg[N\,A^0_4({1}_q,3_g,4_g,\hat{2}_{\bar{q}})
+N\,\,A^0_4({1}_q,4_g,3_g,\hat{2}_{\bar{q}})\nonumber\\
&&\qquad\qquad
-\frac{1}{N}\,\tilde{A}^0_4({1}_q,3_g,4_g,\hat{2}_{\bar{q}})\bigg]\,,\\
\left|M^0_{q,qq\bar{q}}\right|^2+\left|M^0_{q,qq'\bar{q}'}\right|^2&=&e_q^2\,N_{3,q}\,
\bigg[N_F\,B^0_4({{1}_q,3_{q'},4_{\bar{q}'},\hat{2}_{\bar{q}}})\nonumber\\
&&
-\frac{1}{N}\big(C^0_4({{1}_q,3_q,4_{\bar{q}},\hat{2}_{\bar{q}}})+
                  C^0_4({{1}_q,3_q,\hat{2}_{\bar{q}},4_{\bar{q}}})
\nonumber\\&&\qquad
+
                  C^0_4({4_{\bar{q}},\hat{2}_{\bar{q}},{1}_q,3_q})+
                  C^0_4(\hat{2}_{\bar{q}},4_{\bar{q}},{{1}_q,3_q})\big)
\bigg]\nonumber\\
&&+\sum_{q'}e_{q'}^2\,N_{3,q}\,B^0_4({{3}_{q'},1_{q},\hat{2}_{\bar{q}},
{4}_{\bar{q}'}})
\,,\\
\left|M^0_{g,q\bar{q}}\right|^2&=&\sum_{q}e_q^2\,N_{2,g}\,\,A^0_4(1_q,\hat{3}_g,2_{\bar{q}})\,,\\
\left|M^0_{g,gq\bar{q}}\right|^2&=&\sum_{q}e_q^2\,N_{3,g}\,\bigg[N\,A^0_4(1_q,\hat{3}_g,4_g,2_{\bar{q}})
+N\,A^0_4(1_q,4_g,\hat{3}_g,2_{\bar{q}})
\nonumber\\&&\qquad\qquad
-\frac{1}{N}\,\tilde{A}^0_4(1_q,\hat{3}_g,4_g,2_{\bar{q}})\bigg]\,,
\end{eqnarray}
where
\begin{equation}
N_{n,q}=4\,\pi\alpha\,(g^2)^{n-2}\frac{N^2-1}{N}\,2\,(1-\epsilon)\,Q^2\,,
\end{equation}
and
\begin{equation}
N_{n,g}=4\,\pi\alpha\,(g^2)^{n-2}\,2\,Q^2\,.
\end{equation}
All the antenna functions used here 
 can be obtained from explicit expressions for the final-final 
antennae $X_3^0$ and $X_4^0$ in~\cite{Gehrmann-DeRidder:2005cm} by 
crossing in each case the particle denoted by $\hat{n}_i$.

The subtraction terms for both quark and gluon initiated processes are a
combination of final-final and initial-final subtractions. We split the
quark-induced 
 contributions into three terms: quark-plus-two-gluon final states 
at leading and subleading colour,  $\d\hat{\sigma}^{S}_{q,A}$ and 
$\displaystyle \d\hat{\sigma}^{S}_{q,{\tilde{A}}}$ and quark-quark-antiquark final state
$\d\hat{\sigma}^{S}_{q,B}$. Identical-only quark contributions to 
the matrix elements, involving the antenna 
$C^0_4$, have no single collinear limits so they do not need to be
subtracted.  Gluon-induced 
contributions  can also be split into leading and subleading colour,
$\d\hat{\sigma}^{S}_{g,A}$ and $\displaystyle \d\hat{\sigma}^{S}_{g,{\tilde{A}}}$.
The explicit expressions for subtraction terms are given by
\begin{eqnarray}
\d\hat{\sigma}^{S}_{q,A}=&&e_q^2\,N_{3,q}\d\Phi_{3}(k_{g_1},k_{g_2},k_q;p_q,q)\frac{N}{2}\,\nonumber\\
               &&\times\left(D^0_{q,g_1g_2}\,A^0_{Q,Gq}\,J^{(2)}_2(K_G,k_q)+
                     D^0_{g_1g_2q}A^0_{q,GQ}\,J^{(2)}_2(K_G,K_Q)\right)\,,\\
\d\hat{\sigma}^{S}_{q,\tilde{A}}=&&-e_q^2\,N_{3,q}\d\Phi_{3}(k_{g_1},k_{g_2},k_q;p_q,q)\frac{1}{2N}\,\nonumber\\
               &&\times\left(A^0_{q,g_1q}\,A^0_{Q,g_2Q}\,J^{(2)}_2(k_{g_2},K_Q)+
                     A^0_{q,g_2q}\,A^0_{Q,g_1,Q}\,J^{(2)}_2(k_{g_1},K_Q)\right)\,,\\
\d\hat{\sigma}^{S}_{q,B}=&&N_{3,q}\d\Phi_{3}(k_{q},k_{q'},k_{\bar{q}'};p_q,q)\,\Bigg[
\sum_{q'}e_{q'}^2\,E^0_{q,q{q}'}\,A^0_{G,Q'\bar{q}'}\,J^{(2)}_2(K_{Q'},k_{\bar{q}'})
\nonumber\\
       &&+\frac{N_F}{2}\,e_q^2\,\left(E^0_{q,q'\bar{q}'}\,A^0_{Q,Gq}\,J^{(2)}_2(K_{G},k_q)
              +E^0_{qq'\bar{q}'}\,A^0_{q,GQ}\,J^{(2)}_2(K_{G},k_q)\right)\Bigg]\,,\\
\d\hat{\sigma}^{S}_{g,A}=&&\sum_{q}e_q^2\,N_{3,g}\d\Phi_{3}(k_{g},k_{q},k_{\bar{q}};p_g,q)\,N\,\nonumber\\
               &&\times\left(D^0_{g,\bar{q}g}\,A^0_{Q,Gq}\,J^{(2)}_2(K_G,k_q)+
                     D^0_{g,g\bar{q}}A^0_{G,q,{\bar{Q}}}\,J^{(2)}_2(k_q,K_{\bar{Q}})\right.\nonumber\\
               &&\quad+\left.D^0_{g,qg}\,A^0_{{\bar{Q}},G\bar{q}}\,J^{(2)}_2(K_G,k_{\bar{q}})+
                 D^0_{g,gq}A^0_{G,Q\bar{q}}\,J^{(2)}_2(K_Q,k_{\bar{q}})\right)\,,\\
\d\hat{\sigma}^{S}_{g,\tilde{A}}=&&-\sum_{q}e_q^2\,N_{3,g}\d\Phi_{3}(k_{g},k_{q},k_{\bar{q}};p_g,q)\frac{1}{N}\,
\nonumber\\
               &&\times\bigg(\frac{1}{2}A^0_{g,q\bar{q}}\,A^0_{Q,gQ}\,J^{(2)}_2(k_{g},K_Q)+
               \frac{1}{2}A^0_{g,q\bar{q}}\,A^0_{{\bar{Q}},g{\bar{Q}}}
                \,J^{(2)}_2(k_{g},K_{\bar{Q}})
\nonumber\\&&
+\,A^0_{qg\bar{q}}A^0_{g,{Q}{\bar{Q}}}\,J^{(2)}_2(K_{Q},K_{\bar{Q}})\bigg)\,.
\end{eqnarray}
In the above, $X_{I,Jk}$ and $X_{k,IJ}$ denote three-parton antenna functions 
with momenta $I,J$ obtained from a phase space mapping. The combination of these
subtraction terms with the real matrix elements containing three partons in the
final state is finite in all soft and collinear limits and can be integrated
numerically over the three-particle phase space.

On the other hand, the analytical integration of the subtraction terms over the factorized phase space can
be carried out using the results
of Section \ref{sec:antenna-if}. 
For the poles of the integrated terms, we obtain:
\begin{eqnarray}
\d\hat{\sigma}^{S}_{q,A}+\d\hat{\sigma}^{S}_{q,\tilde{A}}+\d\hat{\sigma}^{S}_{q,B}
&=&-2\,\frac{\alpha_s}{2\pi}\Big[
N\,\left({\bf I}^{(1)}_{qg}(t)+{\bf I}^{(1)}_{qg}(s)\right)
-\frac{1}{N}\,{\bf I}^{(1)}_{qq}(u)\nonumber\\
&&\qquad\quad+N_F\,\left({\bf I}^{(1)}_{qg,F}(t)+{\bf I}^{(1)}_{qg,F}(s)\right)\Big]\,\d\hat{\sigma}^{B}_{q}(p,q)\nonumber\\
&&-\frac{\alpha_s}{2\pi}\,\int\frac{dx}{x}\frac{1}{\epsilon}\,
C_F\,p^{(0)}_{qq}(x)\,\d\hat{\sigma}^{B}_{q}(xp,q)\nonumber\\
&&-\frac{\alpha_s}{2\pi}
\,\int\frac{dx}{x}\frac{1}{\epsilon}\,C_F\,p^{(0)}_{gq}(x)\,
\d\hat{\sigma}^{B}_{g}(xp,q)+{\cal O}(\epsilon^0)\,, \\
\d\hat{\sigma}^{S}_{g,A}+\d\hat{\sigma}^{S}_{g,\tilde{A}}&=&
-2\,\frac{\alpha_s}{2\pi}\Big[
N\,\left({\bf I}^{(1)}_{qg}(t)+{\bf I}^{(1)}_{qg}(u)\right)
-\frac{1}{N}\,{\bf I}^{(1)}_{qq}(s)\nonumber\\
&&\qquad\quad+N_F\,\left({\bf I}^{(1)}_{qg,F}(t)+{\bf I}^{(1)}_{qg,F}(u)\right)
\Big]\,\d\hat{\sigma}^{B}_{g}(p,q)\nonumber\\
&&
-\frac{\alpha_s}{2\pi}\,\int\frac{dx}{x}\frac{1}{\eps}\,\left(
N\,p^{(0)}_{gg}(x)+N_F\,p^{(0)}_{gg,F}(x)\right)\,\,\d\hat{\sigma}^{B}_{g}(xp,q)\nonumber\\
&&-\frac{\alpha_s}{2\pi}\,\int\frac{dx}{x}\frac{1}{\eps}\,T_F\,p^{(0)}_{qg}(x)
\,\left(\d\hat{\sigma}^{B}_{q}(xp,q)+\d\hat{\sigma}^{B}_{\bar{q}}(xp,q)\right)
+{\cal O}(\epsilon^0)\,, \nonumber \\
\end{eqnarray}
where $C_F=(N^2-1)/(2N)$, $T_F=\frac{1}{2}$ and the Born cross sections are given by
\begin{eqnarray}
\d\hat{\sigma}^{B}_{q}(p,q)&=&\d\Phi_{2}(k_{g},k_{q};p,q)\,\left|M^0_{q,gq}\right|^2\,J^{(2)}_{2}(k_{g},k_{q}),\\
\d\hat{\sigma}^{B}_{g}(p,q)&=&\d\Phi_{2}(k_{q},k_{\bar{q}};p,q)\,\left|M^0_{g,q\bar{q}}\right|^2\,J^{(2)}_{2}(k_{q},k_{\bar{q}}).
\end{eqnarray}
The poles contained in the operators ${\bf I}^{(1)}_{ij}$ match exactly the ones appearing, with opposite
sign, in the interference of the renormalized one loop amplitudes with the
Born ones. The remaining poles correspond to the mass factorization contributions. Thus, combining
the integrated subtraction terms with the virtual contributions and the mass factorization counterterms,
we obtain a finite contribution, free on any poles in $\epsilon$ that can be integrated over
the two partons phase space.

\subsection{Vector-boson-plus-jet production in hadronic colliders}
The second example we will consider is the production of a vector boson $V$
($V=\gamma$, $Z$, $W^{\pm}$) 
plus a hadronic jet in a hadronic collision. This process is mediated by 
the scattering of two partons into the vector boson and one hard parton. 
The cross section is 
given by
\begin{eqnarray}
d\sigma=&&\int \frac{d\xi_1}{\xi_1}\frac{d\xi_2}{\xi_2}\,
\Big\{%
\sum_{i,j}\left[f_{q_{i}}(\xi_1)f_{\bar{q}_{j}}(\xi_2)+
f_{\bar{q}_{i}}(\xi_1)f_{q_{j}}(\xi_2)\right]\,\d\hat{\sigma}_{q_i\bar{q}_j}
\nonumber\\
&&
\qquad\qquad\quad
+\sum_{i,j}\left[f_{q_{i}}(\xi_1)f_{q_{j}}(\xi_2)+
f_{\bar{q}_{i}}(\xi_1)f_{\bar{q}_{j}}(\xi_2)\right]\,\d\hat{\sigma}_{q_iq_j}
\nonumber\\
&&
\qquad\qquad\quad
+\sum_i \left[(f_{q_{i}}(\xi_1)+f_{\bar{q}_{i}}(\xi_1))\,f_{g}(\xi_2)+
f_{g}(\xi_1)\,(f_{q_{i}}(\xi_2)+f_{\bar{q}_{i}}(\xi_2))
\right]\,\d\hat{\sigma}_{q_ig}
\nonumber\\
&&
\qquad\qquad\quad
+
f_{g}(\xi_1)f_{g}(\xi_2)\,\d\hat{\sigma}_{gg}
\Big\}\,.
\end{eqnarray}
Again we express the partonic cross sections in terms of color ordered
antennae. We first write:
\begin{eqnarray}
\d\hat{\sigma}_{q_i\bar{q}_j}&=&\d\Phi_{2}(k_{g},q;p_q,p_{\bar{q}})
                           \,\left|M^0_{q_i\bar{q}_j,g}\right|^2\,J^{(1)}_{1}(k_{g})\nonumber\\
               &&+\d\Phi_{2}(k_{g},q;p_q,p_{\bar{q}})\,2\Re(M^{1\dagger}_{q_i\bar{q}_j,g}
                 M^0_{q_j\bar{q}_j,g})\,J^{(1)}_{1}(k_{g})\nonumber\\
               &&+\d\Phi_{3}(k_{g_1},k_{g_2},q;p_q,p_{\bar{q}})\,\left|M^0_{q_i\bar{q}_j,gg}\right|^2
                 \,J^{(2)}_{1}(k_{g_1},k_{g_2})\nonumber\\
               &&+\sum_{k,l} \d\Phi_{2}(k_{q},k_{\bar{q}},q;p_{q},p_{\bar{q}})
                  \,\left|M^0_{q_i\bar{q}_j,q_k\bar{q}_l}\right|^2\,J^{(2)}_{1}(k_{q_3},k_{\bar{q}_4})\,,\\
\d\hat{\sigma}_{q_iq_j}&=&\sum_{k,l} \d\Phi_{3}(k_{q_k},k_{q_l},q;p_{q_i},p_{q_j})\,\left|M^0_{q_i{q}_j,q_k{q}_l}\right|^2
               \,J^{(2)}_{1}(k_{q_3},k_{{q}_4})\,,\\
\d\hat{\sigma}_{q_ig}&=&\sum_{j} \d\Phi_{2}(k_{q},q;p_q,p_g)
                           \,\left|M^0_{q_ig,q_j}\right|^2\,J^{(1)}_{1}(k_{q})\nonumber\\
               &&+\sum_{j} \d\Phi_{2}(k_{q},q;p_q,p_g)\,2\Re(M^{1\dagger}_{q_ig,q_j}
                 M^0_{q_ig,q_j})\,J^{(2)}_{1}(k_{q})\nonumber\\
                 &&+\sum_{j} \d\Phi_{3}(k_{q},k_{g},q;p_q,p_g)\,
                   \left|M^0_{q_ig,q_jg}\right|^2\,J^{(2)}_{1}(k_{q},k_{g})\nonumber\\
\d\hat{\sigma}_{gg}&=&\sum_{i,j} \d\Phi_{3}(k_{q},k_{\bar{q}},q;p_{g_1},p_{g_2})\,\left|M^0_{gg,q_i\bar{q}_j}\right|^2
                 \,J^{(2)}_{1}(k_{q},k_{\bar{q}})\,,
\end{eqnarray}
where we have omitted again the momentum arguments of the matrix elements. 
The matrix element for the 
partonic process $ab\rightarrow cdV$ is given by $M_{ab,cd}$ and the 
momentum of the vector boson, appearing in the phase space measure is 
denoted by $q$. 

The real contributions are given by
\begin{eqnarray}
\left|M^0_{q_i\bar{q}_j,g}\right|^2&=&\left|C_{ij}\right|^2\,(v_i^2+a_i^2)
\,N_{1,q\bar{q}}\,A^0_3(\hat{j}_{q},1_g,\hat{i}_{\bar{q}})\label{eq:vbme1}\\
\left|M^0_{q_ig,q_j}\right|^2&=&\left|C_{ij}\right|^2\,(v_i^2+a_i^2)\,N_{1,qg}\,A^0_3(j_{q},\hat{1}_g,\hat{i}_{\bar{q}})\\
\left|M^0_{q_i\bar{q}_j,gg}\right|^2&=&\left|C_{ij}\right|^2\,(v_i^2+a_i^2)\,\frac{1}{2}\,N_{2,q\bar{q}}
 \,\Big[N\,A^0_4(\hat{j}_{q},1_g,2_g,\hat{i}_{\bar{q}})
+N\,\,A^0_4(\hat{j}_{q},2_g,1_g,\hat{i}_{\bar{q}})
\nonumber\\
&&\qquad\qquad\qquad\qquad\qquad-\frac{1}{N}\,
\tilde{A}_4(\hat{j}_{q},1_g,2_g,\hat{i}_{\bar{q}})\Big]\,,\\
\left|M^0_{q_i\bar{q}_j,q_k\bar{q}_l}\right|^2&=&N_{2,q\bar{q}}\,\Big\{
  \delta_{kl}\,\left|C_{ij}\right|^2\,(v_i^2+a_i^2)
  \,B_{4}^{0}(\hat{j}_{q},k_{q},l_{\bar{q}},\hat{i}_{\bar{q}})
\nonumber\\&&\qquad\,\,+
  \delta_{ij}\,\left|C_{kl}\right|^2\,(v_k^2+a_k^2)
  \,B_{4}^{0}(k_{q},\hat{j}_{q},\hat{i}_{\bar{q}},l_{\bar{q}})
\nonumber\\&&\qquad\,\,+
  \delta_{jl}\,\left|C_{ik}\right|^2\,(v_i^2+a_i^2)
  \,B_{4}^{0}(k_{q},\hat{j}_{q},l_{\bar{q}},\hat{i}_{\bar{q}})
\nonumber\\&&\qquad\,\,+
  \delta_{ik}\,\left|C_{jl}\right|^2\,(v_j^2+a_j^2)
  \,B_{4}^{0}(\hat{j}_{q},k_{q},\hat{i}_{\bar{q}},l_{\bar{q}})
\nonumber\\&&\qquad\,\,+
  \delta_{ij}\delta_{kl}\,2\,\Re(C_{ii}C^{\dagger}_{kk})\,
  \left(v_i\,v_k\,
   \hat{B}^0_{4,V}(\hat{j}_{q},k_{q},l_{\bar{q}},\hat{i}_{\bar{q}})
  +a_i\,a_k\,
    \hat{B}^0_{4,A}(\hat{j}_{q},k_{q},l_{\bar{q}},\hat{i}_{\bar{q}})\right)
\nonumber\\&&\qquad\,\,+
  \delta_{ik}\delta_{jl}\,2\,\Re(C_{ii}C^{\dagger}_{jj})\,
  \left(v_i\,v_j\,
  \hat{B}^0_{4,V}(k_{q},\hat{j}_{q},l_{\bar{q}},\hat{i}_{\bar{q}})
  +a_i\,a_j\,
   \hat{B}^0_{4,A}(k_{q},\hat{j}_{q},l_{\bar{q}},\hat{i}_{\bar{q}})\right)
\nonumber\\&&\qquad\,\,+\frac{2}{N}\,
  \delta_{ik}\delta_{kl}\,\left|C_{ij}\right|^2\,(v_i^2+a_i^2)
   \,C_{4}^{0}(\hat{j}_{q},k_{q},l_{\bar{q}},\hat{i}_{\bar{q}})
\nonumber\\&&\qquad\,\,+\frac{2}{N}\,
  \delta_{ij}\delta_{jl}\,\left|C_{ik}\right|^2\,(v_i^2+a_i^2)
   \,C_{4}^{0}(k_{q},\hat{j}_{q},l_{\bar{q}},\hat{i}_{\bar{q}})
\nonumber\\&&\qquad\,\,+\frac{2}{N}\,
  \delta_{jl}\delta_{kl}\,\left|C_{ij}\right|^2\,(v_i^2+a_i^2)
   \,C_{4}^{0}(\hat{i}_{\bar{q}},l_{\bar{q}},k_{q},\hat{j}_{q})
\nonumber\\&&\qquad\,\,+\frac{2}{N}\,
  \delta_{ij}\delta_{ik}\,\left|C_{jk}\right|^2\,(v_j^2+a_j^2)
   \,C_{4}^{0}(l_{\bar{q}},\hat{i}_{\bar{q}},k_{q},\hat{j}_{q})
\Big\}\\
\left|M^0_{q_iq_j,q_kq_l}\right|^2&=&N_{2,qq}\,
  \left(1-\frac{\delta_{kl}}{2}\right)\nonumber\\
&&\times\Big\{
\delta_{jl}\,\left|C_{ik}\right|^2\,(v_i^2+a_i^2)
  \,B_{4}^{0}(k_{q},l_{q},\hat{j}_{\bar{q}},\hat{i}_{\bar{q}})
\nonumber\\&&\quad +
\delta_{ik}\,\left|C_{jl}\right|^2\,(v_j^2+a_j^2)
  \,B_{4}^{0}(l_{q},k_{q},\hat{i}_{\bar{q}},\hat{j}_{\bar{q}})
\nonumber\\&&\quad+
  \delta_{ik}\delta_{jl}\,2\,\Re(C_{ii}C^{\dagger}_{jj})\,
  \left(v_i\,v_j\,
   \hat{B}^0_{4,V}(k_{q},l_{q},\hat{j}_{\bar{q}},\hat{i}_{\bar{q}})
  +a_i\,a_j\,
   \hat{B}^0_{4,A}(k_{q},l_{q},\hat{j}_{\bar{q}},\hat{i}_{\bar{q}})\right)
\Big\}\nonumber\\
&&+N_{2,qq}\,
\left(\delta_{ij}\,\left(1-\delta_{kl}\right)+\frac{\delta_{kl}}{2}\right)
\nonumber\\
&&\times\Big\{
\delta_{jk}\,\left|C_{il}\right|^2\,(v_i^2+a_i^2)
  \,B_{4}^{0}(l_{q},k_{q},\hat{j}_{\bar{q}},\hat{i}_{\bar{q}})
\nonumber\\
&&\quad+
\delta_{il}\,\left|C_{jk}\right|^2\,(v_j^2+a_j^2)
 \,B_{4}^{0}(k_{q},l_{q},\hat{i}_{\bar{q}},\hat{j}_{\bar{q}})
\nonumber\\
&&\quad+
  \delta_{jk}\delta_{il}\,2\,\Re(C_{ii}C^{\dagger}_{jj})\,
  \left(v_i\,v_j
   \,\hat{B}^0_{4,V}(l_{q},k_{q},\hat{j}_{\bar{q}},\hat{i}_{\bar{q}})
  +a_i\,a_j\,
  \hat{B}^0_{4,A}(l_{q},k_{q},\hat{j}_{\bar{q}},\hat{i}_{\bar{q}})\right)
\nonumber\\
&&\quad-
\frac{2}{N}\,\delta_{jl}\delta_{jk}\,\left|C_{ij}\right|^2\,(v_i^2+a_i^2)
  \,C_{4}^{0}(\hat{i}_{\bar{q}},\hat{j}_{\bar{q}},k_{q},l_{q})
\nonumber\\
&&\quad-
\frac{2}{N}\,
  \delta_{il}\delta_{jl}\,\left|C_{ik}\right|^2\,(v_i^2+a_i^2)
  \,C_{4}^{0}(k_{q},l_{q},\hat{i}_{\bar{q}},\hat{j}_{\bar{q}})
\nonumber\\
&&\quad-
\frac{2}{N}\,
  \delta_{ik}\delta_{jk}\,\left|C_{il}\right|^2\,(v_i^2+a_i^2)
  \,C_{4}^{0}(l_{q},k_{q},\hat{i}_{\bar{q}},\hat{j}_{\bar{q}})
\nonumber\\
&&\quad-
\frac{2}{N}\,\delta_{ik}\delta_{il}\,\left|C_{ij}\right|^2\,(v_j^2+a_j^2)
  \,C_{4}^{0}(\hat{j}_{\bar{q}},\hat{i}_{\bar{q}},k_{q},l_{q})
\Big\}\,,\\
\left|M^0_{g_1q_i,g_2{q}_j}\right|^2&=&\left|C_{ij}\right|^2\,(v_i^2+a_i^2)\,
\,N_{2,qg}
 \,\Big[N\,A^0_4(j_{q},\hat{1}_g,2_g,\hat{i}_{\bar{q}})+
  N\,A^0_4(j_{q},2_g,\hat{1}_g,\hat{i}_{\bar{q}})
\nonumber\\
&&\qquad\qquad\qquad\qquad\qquad-\frac{1}{N}\,
\tilde{A}_4(j_{q},\hat{1}_g,2_g,\hat{i}_{\bar{q}})\Big]\,,\\
\left|M^0_{g_1g_2,q_i\bar{q}_j}\right|^2&=&\left|C_{ij}\right|^2
  \,(v_i^2+a_i^2)\,N_{2,gg}
 \,\Big[N\,A^0_4(i_{q},\hat{1}_g,\hat{2}_g,j_{\bar{q}})
  +N\,A^0_4(i_{q},\hat{2}_g,\hat{1}_g,j_{\bar{q}})
\nonumber\\
&&\qquad\qquad\qquad\qquad\qquad-\frac{1}{N}\,
\tilde{A}_4(i_{q},\hat{1}_g,\hat{2}_g,j_{\bar{q}})\Big]\,,\label{eq:vbmef}
\end{eqnarray}
where
\begin{eqnarray}
N_{n,q\bar{q}}&=&N_{n,qq}=4\,\pi\alpha_V\,(g^2)^{n-2}\frac{N^2-1}{N^2}\,(1-\epsilon)\,Q^2\,,\\
N_{n,qg}&=&4\,\pi\alpha_V\,(g^2)^{n-2}\frac{1}{N}\,Q^2\,,\\
N_{n,qg}&=&4\,\pi\alpha_V\,(g^2)^{n-2}\frac{1}{N^2-1}\frac{1}{1-\epsilon}\,Q^2\,,
\end{eqnarray}
with the coupling constants given by
\begin{equation}
\alpha_\gamma=\alpha=\frac{e^2}{4\pi}\,,\qquad \alpha_{W}=\frac{G_F\,M_W^2\sqrt{2}}{4\pi}\,,\qquad 
\alpha_{Z}=\frac{G_F\,M_Z^2\sqrt{2}}{64\pi}\,.
\end{equation}
The vector and axial couplings of the quarks to the vector bosons are
\begin{eqnarray}
&&v_u^\gamma=\frac{2}{3}\,,\qquad v_d^\gamma=-\frac{1}{3}\,,\qquad a_u^\gamma=a_d^\gamma=0\,,\nonumber\\
&&v_u^Z=1-\frac{8}{3}\sin^2\theta_W\,,\qquad v_d^Z=-1+\frac{4}{3}\sin^2\theta_W\,,\qquad 
a_u^Z=-1\,,\qquad a_d^Z=1\,,\nonumber\\
&&v_u^W=v_d^W=\frac{1}{\sqrt{2}}\,,\qquad a_u^W=a_d^W=-\frac{1}{\sqrt{2}}\,.
\end{eqnarray}
The flavor mixing matrices, $C_{ij}$ are given by $\delta_{ij}$ in the case of $\gamma$ and
$Z$ production and by the CKM matrix in case of $W$ production.
Finally, the colour ordered antenna functions appearing in 
eqs. (\ref{eq:vbme1}) to (\ref{eq:vbmef})
can be obtained from explicit expressions for the final-final 
antennae $X_3^0$ and $X_4^0$ in~\cite{Gehrmann-DeRidder:2005cm} by 
crossing the particles denoted with hats. 

The subtraction terms for this process involve initial-initial and 
initial-final antennae as there are at most two partons in the final state,
final-final antennae are not needed. Only antennae $A^0_4$ and $B^0_4$ 
contain singular configurations and, thus, require to be subtracted. We
find the following subtraction terms, classified according to the partonic
reaction they must be combined with,
\begin{eqnarray}
\d\hat{\sigma}^S_{q_i\bar{q}_j,gg}=&&\left|C_{ij}\right|^2\,(v_i^2+a_i^2)\,\frac{1}{2}\,N_{2,q\bar{q}}\,
   \d\Phi_{3}(k_{g_1},k_{g_2},q;p_q,p_{\bar{q}})
\nonumber\\
&&\times\Big\{
   N\left[D^0_{q_i,g_1g_2}\,A^0_{Q_i\bar{q}_j,G}\,J^{(1)}_{1}(K_{G})
         +D^0_{\bar{q}_j,g_1g_2}\,A^0_{q_i\bar{Q}_j,G}\,J^{(1)}_{1}(K_{G})\right]
\nonumber\\&&\quad
  -\frac{1}{N}\left[A^0_{q_i\bar{q}_j,g_1}\,A^0_{Q_i\bar{Q}_j,G_2}\,J^{(1)}_{1}(K_{G_2})
                   +A^0_{q_i\bar{q}_j,g_2}\,A^0_{Q_i\bar{Q}_j,G_1}\,J^{(1)}_{1}(K_{G_1})\right]
\Big\}\,,\\
\d\hat{\sigma}^S_{q_i\bar{q}_j,q_k\bar{q}_l}=&&\sum_{k,l}\,N_{2,q\bar{q}}\,
   \d\Phi_{3}(k_{q},k_{\bar{q}},q;p_{q},p_{\bar{q}})
\nonumber\\
&&\times\Big\{
  \delta_{kl}\,\left|C_{ij}\right|^2\,(v_i^2+a_i^2)\,\frac{1}{2}\,
\Big(E^0_{q_i,q'_k\bar{q}'_l}\,A^0_{Q_i\bar{q}_j,G}\,J^{(1)}_{1}(K_{G})
\nonumber\\
&&\qquad\qquad\qquad\qquad\qquad\quad
+E^0_{\bar{q}_j,q'_k\bar{q}'_l}\,A^0_{q_i\bar{Q}_j,G}\,J^{(1)}_{1}(K_{G})\Big)
\nonumber\\&&\quad\,\,+
  \delta_{jl}\,\left|C_{ik}\right|^2\,(v_i^2+a_i^2)\,E^0_{\bar{q}_{j}q'_i,\bar{q}_l}\,\,A^0_{Q_iG,Q_k} \,J^{(1)}_{1}(K_{Q_k})
\nonumber\\&&\quad\,\,+
  \delta_{ik}\,\left|C_{jl}\right|^2\,(v_j^2+a_j^2)\,E^0_{{q}_{i}\bar{q}'_j,{q}_k}\,\,A^0_{\bar{Q}_jG,\bar{Q}_l} \,J^{(1)}_{1}(K_{Q_l})
\Big\}\,,\\
\d\hat{\sigma}^S_{q_iq_j,q_kq_l}=&&\sum_{k,l}\,N_{2,qq}\,
   \d\Phi_{3}(k_{q_k},k_{q_l},q;p_{q_i},p_{q_j}) 
\nonumber\\
&&\times\Big\{
\left(1-\frac{\delta_{kl}}{2}\right)\delta_{jl}\,\left|C_{ik}\right|^2\,(v_i^2+a_i^2)
E^0_{q_j,q_lq'_k}\,A^0_{q_iG,Q_k}\,J^{(1)}_{1}(K_{Q_k})
\nonumber\\&&\quad\,\,+
\left(1-\frac{\delta_{kl}}{2}\right)\delta_{ik}\,\left|C_{jl}\right|^2\,(v_j^2+a_j^2)
E^0_{q_i,q_kq'_l}\,A^0_{q_jG,Q_l}\,J^{(1)}_{1}(K_{Q_l})
\nonumber\\&&\quad\,\,+
\left(\delta_{ij}\,\left(1-\delta_{kl}\right)+\frac{\delta_{kl}}{2}\right)\,
\delta_{jk}\,\left|C_{il}\right|^2\,(v_i^2+a_i^2)\,E^0_{q_j,q_kq'_l}\,A^0_{q_iG,Q_l}
\,J^{(1)}_{1}(K_{Q_l})
\nonumber\\&&\quad\,\,+
\left(\delta_{ij}\,\left(1-\delta_{kl}\right)+\frac{\delta_{kl}}{2}\right)\,
\delta_{il}\,\left|C_{jk}\right|^2\,(v_j^2+a_j^2)\,E^0_{q_i,q_lq'_k}\,A^0_{q_jG,Q_k}
\,J^{(1)}_{1}(K_{Q_k})
\Big\}\,,
\nonumber \\
\\
\d\hat{\sigma}^S_{q_ig,{q}_jg}=&&\sum_{j}\left|C_{ij}\right|^2\,(v_i^2+a_i^2)\,N_{2,qg}\,
   \d\Phi_{3}(k_{q},k_{g},q;p_q,p_{g})
\nonumber\\
&&\times\Big\{
   N\,\Big[D^0_{q_ig,g}\,A^0_{Q_iG,Q_j}\,J^{(1)}_{1}(K_{Q_j})
         +D^0_{g,gq_j}\,A^0_{q_iG,Q_j}\,J^{(1)}_{1}(K_{Q_j})
\nonumber\\
&&\qquad\quad         +D^0_{g,q_jg}\,A^0_{q_i\bar{Q}_j,G}\,J^{(1)}_{1}(K_{G})\Big]
\nonumber\\&&\quad
  -\frac{1}{N}\left[A^0_{q_i,gq_j}\,A^0_{Q_ig,Q_j}\,J^{(1)}_{1}(K_{Q_j})
+A^0_{q_ig,q_j}\,A^0_{Q_i\bar{Q}_j,G}\,J^{(1)}_{1}(K_{G})\right]
\Big\}\,,\\
\d\hat{\sigma}^S_{g_1g_2,q_i\bar{q}_j}=&&\sum_{i,j}\left|C_{ij}\right|^2\,(v_i^2+a_i^2)\,N_{2,gg}\,
   \d\Phi_{3}(k_{q},k_{\bar{q}},q;p_{g_1},p_{g_2})
\nonumber\\
&&\times
   \frac{N^2-1}{N}\left[D^0_{g_1g_2,q_i}\,A^0_{\bar{Q}_iG,\bar{Q}_j}\,J^{(1)}_{1}(K_{Q_j})
         +D^0_{g_1g_2,\bar{q}_j}\,A^0_{Q_jG,Q_i}\,J^{(1)}_{1}(K_{Q_i})\right]\,,
\end{eqnarray}
Antennae of the form $X_{iJ,K}$ and $X_{IJ,K}$ correspond to antennae where the momenta of the 
particles denoted with capital letters are obtained by initial-final and initial-initial 
phase space mappings respectively.

The integrated form of the subtraction terms can be obtained inmediately using the results in Sections
 \ref{sec:antenna-if} and \ref{sec:antenna-ii}. The singular pieces of these terms are then given by
\begin{eqnarray}
\d\hat{\sigma}^S_{q_i\bar{q}_j,gg}=&&-2\frac{\alpha_s}{2\pi}\Big[
   N\,\left({\bf I}^{(1)}_{qg}(t)+{\bf I}^{(1)}_{qg}(u)\right)-
   \frac{1}{N}\,{\bf I}^{(1)}_{qq}(s)\Big]\,\d\hat{\sigma}^B_{q_i\bar{q}_j}(p_q,p_{\bar{q}})\nonumber\\
&&-\frac{\alpha_s}{2\pi}\int\frac{dx}{x}\frac{1}{\epsilon}\,C_F\,p^{(0)}_{qq}(x)\,
\left(\d\hat{\sigma}^B_{q_i\bar{q}_j}(x\,p_q,p_{\bar{q}})+\d\hat{\sigma}^B_{q_i\bar{q}_j}(p_q,x\,p_{\bar{q}})\right)
+{\cal O}(\epsilon^0)\,,\\
\d\hat{\sigma}^S_{q_i\bar{q}_j,q_k\bar{q}_l}=&&-2\frac{\alpha_s}{2\pi}\Big[
   N_F\,\left({\bf I}^{(1)}_{qg,F}(t)+{\bf I}^{(1)}_{qg,F}(u)\right)\Big]
\,\d\hat{\sigma}^B_{q_i\bar{q}_j}(p_q,p_{\bar{q}})\nonumber\\
&&-\frac{\alpha_s}{2\pi}\int\frac{dx}{x}\frac{1}{\epsilon}\,C_F\,p^{(0)}_{gq}(x)\,
\left(\d\hat{\sigma}^B_{q_ig}(p_q,x\,p_g)+\d\hat{\sigma}^B_{\bar{q}_jg}(p_{\bar{q}},x\,p_g)\right)
+{\cal O}(\epsilon^0)\,,\\
\d\hat{\sigma}^S_{q_iq_j,q_kq_l}=&&
-\frac{\alpha_s}{2\pi}\int\frac{dx}{x}\frac{1}{\epsilon}\,C_F\,p^{(0)}_{gq}(x)\,
\left(\d\hat{\sigma}^B_{q_ig}(p_q,x\,p_g)+\d\hat{\sigma}^B_{{q}_jg}(p_{\bar{q}},x\,p_g)\right)
+{\cal O}(\epsilon^0)\,,\\
\d\hat{\sigma}^S_{q_ig,q_jg}=&&-2\frac{\alpha_s}{2\pi}\Big[
   N\,\left({\bf I}^{(1)}_{qg}(u)+{\bf I}^{(1)}_{qg}(s)\right)-
   \frac{1}{N}\,{\bf I}^{(1)}_{q\bar{q}}(t)\nonumber\\
&&\qquad\quad+N_F\,\left({\bf I}^{(1)}_{qg,F}(u)+{\bf I}^{(1)}_{qg,F}(s)\right)
\Big]\,\d\hat{\sigma}^B_{q_ig}(p_q,p_g)\nonumber\\
&&-\frac{\alpha_s}{2\pi}\int\frac{dx}{x}\frac{1}{\epsilon}\,C_F\,p^{(0)}_{qq}(x)\,
\d\hat{\sigma}^B_{q_ig}(x\,p_q,p_{g})\nonumber\\
&&-\frac{\alpha_s}{2\pi}\int\frac{dx}{x}\frac{1}{\epsilon}\,\left(N\,p^{(0)}_{gg}(x)+N\,p^{(0)}_{gg,F}(x)\right)\,
\d\hat{\sigma}^B_{q_ig}(p_q,x\,p_{g})\nonumber\\
&&-\frac{\alpha_s}{2\pi}\int\frac{dx}{x}\frac{1}{\epsilon}\,T_F\,p^{(0)}_{qg}(x)\,
\sum_j \d\hat{\sigma}^B_{q_i\bar{q}_j}(p_q,x\,p_{\bar{q}})
+{\cal O}(\epsilon^0)\,,\\
\d\hat{\sigma}^S_{gg,q_i\bar{q}_j}=
&&-\frac{\alpha_s}{2\pi}\int\frac{dx}{x}\frac{1}{\epsilon}\,2\,T_F\,p^{(0)}_{qg}(x)\,
\sum_i\left( \d\hat{\sigma}^B_{q_ig}(x\,p_q,p_g)+\d\hat{\sigma}^B_{\bar{q}_ig}(x\,p_{\bar{q}},p_{g})\right)
+{\cal O}(\epsilon^0)\,,\nonumber \\
\end{eqnarray}
where the Born cross sections are given by
\begin{eqnarray}
\d\hat{\sigma}^B_{q_i\bar{q}_j}&=&\d\Phi_{2}(k_{g},q;p_q,p_{\bar{q}})
                           \,\left|M^0_{q_i\bar{q}_j,g}\right|^2\,J^{(1)}_{1}(k_{g})\,,\\
\d\hat{\sigma}^B_{q_ig}&=&\sum_{j} \d\Phi_{2}(k_{q},q;p_q,p_g)
                           \,\left|M^0_{q_ig,q_j}\right|^2\,J^{(1)}_{1}(k_{q})\,.
\end{eqnarray}
Again, the poles in the operators ${\bf I}^{(1)}_{ij}$ are canceled by the virtual contributions 
whereas the ones associated to the Altarelli-Parisi kernels drop out once combined with the mass
factorization counterterms.

\section{Conclusions and Outlook}
\label{sec:conc}
In this paper, we have generalized the antenna subtraction method for the 
calculation of higher order QCD
corrections to exclusive collider observables to situations with partons in 
the initial state. 

The basic ingredients to the subtraction terms, the antenna 
functions, can be 
 obtained from the known final-state antenna functions by 
simple crossing. We derived the factorization of an multi-parton phase space 
into an antenna phase space (required for the analytic integration of 
the subtraction terms) and a reduced phase space of lower multiplicity,
for antennae with one or two hard radiator partons in the initial state
(initial-final and initial-initial antennae).   
Explicit phase space factorization and parameterization formulae were presented 
for NLO and NNLO calculations. We derived all integrated 
initial-final and initial-initial antennae relevant at NLO, and demonstrated 
their application on two example calculations. 

A major advantage of the antenna subtraction method is its straightforward 
extension to NNLO calculations. Our results are a significant 
step towards NNLO calculations of hadron collider observables. Using the 
phase space factorizations presented here, NNLO subtraction terms for 
jet production observables at hadron colliders can be constructed from 
known building blocks. Their analytic 
integration over the antenna phase spaces relevant to NNLO calculations is 
still an outstanding issue. It is however anticipated that usage of 
techniques similar to those applied for the 
integration of the final-final antennae will help to perform these integrals 
in a systematic and efficient way.

First applications of the method presented here, once the corresponding NNLO integrated
antennae are available, could be NNLO calculations of 
two-jet production or vector-boson-plus-jet production at hadron colliders,
and of two-plus-one-jet production in deep inelastic scattering. Further 
extensions of the method could include 
radiation off massive particles, thus allowing the NNLO calculation of 
top quark pair production at hadron colliders.

Another important extension of subtraction methods is the combination with 
parton shower algorithms~\cite{mcnlo}, thus allowing for a full partonic 
event generation to NLO accuracy. This task was fully 
accomplished so far only for 
one  QCD subtraction method~\cite{frixione}. While fixed order NLO 
calculations are independent of the subtraction method used, there can be 
a residual
dependence on the method in matched NLO-plus-parton-shower calculations, 
since unintegrated and integrated subtraction terms are treated differently 
in the parton shower. With the formulation of the antenna subtraction 
method for initial state radiation presented here, it will become possible 
to construct antenna-based parton showers for hadronic collisions.

\acknowledgments
TG wishes to thank S.~Jadach for interesting discussions on extensions of the antenna subtraction 
method. 
This research was supported by the Swiss National Science Foundation
(SNF) under contract 200020-109162. AD would like to thank 
The Austrian Federal Ministry for Education, Science and Culture, 
The High Energy Physics Institute of the Austrian Academy of Sciences,
The Erwin Schr\"odinger International Institute of Mathematical Physics
and Vienna Convention Bureau for economical support during the 
Vienna Central European Seminar on Particle Physics and Quantum Field
Theory where part of this work was presented.

\end{document}